\documentclass[manuscript]{aastex}

\newcommand{\eg}{{\it e.g.}}
\newcommand{\etal}{{\it et al.}}
\newcommand{\ie}{{\it i.e.}}
\newcommand{\viz}{{\it viz.}}
\newcommand{\Rsun}{\ensuremath{R_{\odot}}}
\newcommand{\Msun}{\ensuremath{M_{\odot}}}
\newcommand{\mdot}{\ensuremath{\dot{M}}}
\newcommand{\mdotten}{\ensuremath{\dot{M}_{-10}}}
\newcommand{\vsqav}{\ensuremath{v_k^2}}
\newcommand{\lum}{\ensuremath{L_{36}}}

\shorttitle{Collective Properties of HMXBs}
\shortauthors{Bhadkamkar and Ghosh}

\begin{document}


\title{Collective Properties of \\ X-ray Binary Populations of Galaxies.I.
 \\ Luminosity and Orbital Period Distributions of \\ High-Mass X-ray Binaries}


\author{Harshal Bhadkamkar and Pranab Ghosh}
\affil{Department of Astronomy \& Astrophysics, Tata Institute of
  Fundamental Research, Mumbai 400 005, India}



\begin{abstract}
We introduce a method for obtaining the X-ray luminosity function
(XLF) and the binary-period distribution of populations of high-mass
X-ray binaries (HMXBs) in the stellar fields (\ie, outside globular
clusters) of normal galaxies. We start from standard distributions of 
the parameters of those primordial binaries which are the progenitors of
HMXBs, and follow the transformation of these ditributions with
the aid of a Jacobian formalism as the former evolve into the latter
through the processes of the first mass transfer and the supernova
(SN) that follows. We discuss the distributions of the post-SN
binaries and the HMXBs. We show that our calculated model XLF has
a differential slope $\approx -1.6$ with a flattening at low
luminosities, in excellent agreement with observations. The calculated
binary-period distribution, which basically has a slightly sloping
plateau-like character at intermediate periods, with a rise to this
plateau at shorter periods and fall-off from it at longer periods, is  
in agreement with the observed distribution within observational 
uncertainties. We discuss the physical origin of these distributions.
We demonstrate that, while the effects of both (a) the distribution 
of the properties of the massive companion in the HMXBs, and (b) the 
primordial orbital distribution and the SN dynamics are important, 
the former appear to be dominant in determining the XLF, and the 
latter in determining the HMXB binary-period distribution. We discuss 
the possible roles of stellar-mass black holes and ultra-luminous 
X-ray sources (ULX) in the observed ``universal'' XLF of HMXBs. 
\end{abstract}





\section{Introduction}
\label{sec:intro}

Four decades have passed now since the {\it Uhuru} discovery of the
X-ray binary (henceforth XRB) Cen X-3 was first reported
\citep{giet71}. What began in the early- to mid-1970s as detailed 
individual studies of the small number of XRBs known at the time,
as also pioneering works on the construction of general models of 
such binaries \citep{Lambet1,PriRe,DavOs} has now blossomed into a 
mature, rich subject (for overviews, see, \eg, Shapiro \& Teukolsky 
1983; Ghosh 2007), with a large number of XRBs discovered 
and catalogued today in the Milky Way, the Magellanic Clouds,
and nearby external galaxies, and the fine details of the behavior 
of each class of XRBs well-recorded from timing and spectral studies 
with the aid of several generations of X-ray observatories of 
ever-increasing sensitivity and resolution. It is thus becoming 
possible now to do studies of the {\it collective} properties of 
XRB populations, obtaining statistically meaningful distributions 
of some of their essential properties \citep{KimFab04,Grimetal02,
Grimetal03,Gil04,GGS04a,GGS04b,KimFab10,LPH05,LPH06},
and exploring the essential physics underlying these distributions, 
which is rooted in the basic scenarios for the formation, 
evolution, and dynamics of XRBs (van den Heuvel 1983, 1991,
1992, 2001), which have been constructed 
gradually over these four decades, and which are widely accepted now.     
By collective properties, we mean here the distributions of X-ray 
luminosities and binary periods of these XRBs, the distribution
of pulse periods of those binaries which exhibit periodic X-ray pulses,   
and perhaps also the distribution of their spectral properties at a 
future date. 

In this series of papers, we explore the theoretical underpinnings of
the observed distribution of the collective properties of XRBs in
normal galaxies. We focus here on XRBs 
{\it outside} the globular clusters of these galaxies (\ie, in the stellar 
field of the galaxy), where the effects of encounters 
between XRBs and neighboring stars, as well as those between XRBs
themselves, are generally thought to be negligible, due to the 
relatively low stellar density there. This situation is at the opposite
limit to that obtaining in dense cores of globular clusters, where
such encounters are thought to be {\it dominant} in determining the
formation, evolution, and destruction of XRBs, and so their collective properties. 
We have studied this latter limit of XRBs in globular 
clusters in an earlier series of papers (Banerjee \& Ghosh 2006, 2007,
2008). The current study may thus be regarded as complementary to the 
earlier one. We shall confine ourselves here to the distributions of 
X-ray luminosity and binary orbital period, since the most detailed 
distributions available are on these properties, and also
since these properties are most readily associated with binary evolutionary 
characteristics. To understand the distributions of properties like pulse
periods of X-ray pulsars and spectral parameters of XRBs, one must also take 
into account the details of the accretion torque on the neutron star, 
and of the X-ray emission processes. We defer these projects to the
future, noting that with $\sim$ 140 X-ray pulsars already known, the 
pulse-period distribution may be amenable to such a future study. 

The crucial simplification that makes a straightforward approach 
feasible in the limit that we study here is that
the absence of any significant influence on a given XRB of either the 
stellar background or other XRBs implies that each XRB can be thought
of as evolving in {\it isolation}, following the well-studied and 
now widely-accepted scenarios for the formation and evolution 
of individual XRBs, starting from primordial binaries 
generated in normal star-formation activity in a galaxy (see, \eg,
van den Heuvel 2001 and the references therein). 
This, in turn, implies that we need only have a knowledge of the 
distribution of the essential parameters of these primordial 
binaries, and evolve these distributions through the essential 
processes that occur during the evolution of a primordial binary
into a XRB. The task becomes even simpler when only initial and
final states matter in the above processes, since we need only 
keep track of these states to carry out the transformation of 
the relevant probability distribution, and so connect the primordial
distribution to the XRB distribution in a one-to-one 
correspondence. This is, in fact, the case 
for massive or high-mass X-ray binaries (henceforth HMXBs). On 
the other hand, when following one or more of these evolutionary processes 
requires the knowledge of the entire evolution during a particular
process, and not just of the initial and final states, the 
calculation becomes more laborious (although still
straightforward in principle). This is the case for 
low-mass X-ray binaries (henceforth LMXBs). 

In this first paper of the series, we focus our attention on HMXBs
containing neutron stars.
A particular circumstance that helps in this case is that, because 
of the short evolutionary timescales of the massive companions 
of the neutron stars in HMXBs, the timescale for the entire evolution 
from the primordial binary stage to the HMXB stage (as also the 
operational lifetime of the HMXB) is short \citep{vdH01} compared to the 
timescale on which the star-formation rate (which determines the 
creation rate of primordial binaries) evolves \citep{Madau,GW01}. 
As a consequence, 
the entire process we study here can be viewed as happening at a
constant star-formation rate, so that the star-formation history of
the galaxy adds no extra complication. Figuratively 
speaking, it is as if we are taking a ``snapshot'' of the above
processes confined entirely to one epoch (the current one, 
$z = 0$, for the HMXB populations of the Milky Way and other 
local galaxies, and an appropriately earlier epoch for the HMXB 
populations of spiral galaxies at significant redshifts). 

This argument does not, of course, apply to LMXBs, 
since their evolutionary timescales are comparable to or longer 
than \citep{vdH91} those on which the star-formation rate evolves.
Accordingly, we have to explicitly take into account the 
evolution in the star-formation rate when we apply our scheme 
to LMXBs, as we shall describe in later papers 
in the series. Thus, for example, for 
describing the LMXB populations of local galaxies, we shall 
need to use not only the star-formation rate at $z = 0$ but 
also the rates at $z > 0$, \ie, the star-formation history of
the galaxy.
This difference between the evolutionary characteristics of 
HMXB and LMXB populations is, of course, closely 
related to the earlier statement about the initial-to-final 
state correspondence for HMXBs versus the additional role of       
intermediate states for LMXBs.

In the following sections, we first describe the distributions of the 
essential parameters of the primordial binary populations that we
adopt for our study of HMXBs, following well-established 
scenarios and norms in the literature. In particular, we adopt the 
standard initial mass function (henceforth IMF) and the standard 
log-uniform distribution (also known as \"Opik's law) for the orbital 
separation, as is done in the literature. We then describe how 
these parameters change in the essential processes that occur as
a primordial binary evolves \citep{vdH01}. The first change occurs 
when the primary
evolves, fills its Roche lobe, and transfers its envelope to the 
secondary, leaving behind its He core. The second change occurs when 
this He core finishes its further evolution, and explodes in the
supernova (SN) that creates the neutron star. We give explicit
relations connecting the binary parameters in the initial and final
states of these two processes. Using these relations, we show how
the standard rules of probability transformation enable us to 
derive the distributions of the post-SN binary parameters from the
original primordial binary parameters through the Jacobian 
formalism.  Next, we consider what happens when the massive 
companion to the neutron star evolves off the main sequence, 
becoming a giant/supergiant, and driving a strong stellar wind, 
from which the neutron star accretes matter, thus turning on the 
HMXB phase \citep{vdH01}. 

We adopt standard stellar-wind models from the literature
and show how we can derive the distributions of 
both the X-ray luminosities $L_X$ and the orbital periods 
$P_{orb}$ of the HMXBs from the distributions we obtained above. 
We then compare our calculated distributions with the observed 
distributions of $L_X$ \citep{Grimetal02,Gil04,GGS04a}  
and $P_{orb}$ \citep{LPH05,LPH06} reported in the 
literature. We show that the main feature of the observed
$L_X$-distribution, \viz, the power-law behavior with a 
differential slope $dN/dL_X \approx -1.6$ over a wide range
of $L_X$ \citep{Gil04} is reproduced well in our calculated 
distribution over the luminosity range covered by
neutron-star HMXBs.
At the lowest luminosities, below the range covered in the
above observational work, our calculated distribution shows
a shallower rise in $dN/dL_X$ with decreasing $L_X$. 
We show that this feature is consistent with
recent observations (Shtykovskiy \& Gilfanov 2005a,b) of
the Magellanic Clouds. For the $P_{orb}$-distribution,
we show that the general features of the observed distribution
are in reasonable agreement with the results of our calculations
within observational uncertainties.         

Subsequent sections of the paper are arranged as follows.
In Sec.\ref{sec:hmxboverview}, we give a brief overview of the
formation of neutron-star HMXBs from primordial binaries and their 
subsequent evolution, and we describe the standard distribution 
of primordial binary parameters that we adopt. 
In Sec.\ref{sec:paramchange}, we describe how the  
binary parameters evolve through the processes of (a) the first mass
transfer in the system and 
(b) the subsequent supernova of the He-core of the primary. 
In Sec.\ref{sec:jacob}, we introduce the Jacobian formalism for
calculating how the distribution of binary parameters transforms 
as these parameters evolve as described in the previous section. 
In Sec.\ref{sec:psnpdf}, we discuss the distribution of the essential
parameters of the post-SN binary. In Sec.\ref{sec:psntoxrb}, we 
detail the evolution of the post-SN binary into a HMXB, summarizing 
the essentials of the stellar and stellar-wind models that provide these
details, and indicating the further transformation of distributions
needed at this point. In Sec.\ref{sec:hmxbdist}, we discuss our
calculated HMXB distributions of luminosity (\ie, the X-ray luminosity 
function or XLF) and binary period, compare them with observed 
distributions, and discuss the effect on our model calculations of 
varying essential inputs like the initial mass function (IMF), stellar
wind models, and so on. Finally, in Sec.\ref{sec:discussion}, we discuss
our results from various angles, including issues about primordial 
binary parameters, roles of Be-star binaries and black holes,
physical origins of some essential features of the HMXB distributions,
and then summarize our conclusions and future outlook. Details of some
of our calculational methods are summarized in Appendix A.
             
\section{HMXB Formation and Evolution}
\label{sec:hmxboverview}

Scenarios for the formation and evolution of HMXBs have been studied in
detail since the early 1970s, and a ``standard'' picture has now
emerged, which is very well-documented 
\citep{vdH83,vdH91,vdH92,vdH01}, thus making it
unnecessary to recount it here in detail. Very briefly, one starts with 
a primordial binary of two massive stars. The more massive of
the two (\ie, the primary) evolves faster, ends its main sequence life
and expands rapidly to become a giant, the whole process occurring on
a timescale of $\sim 10^6-10^7$ years. The giant overflows its Roche lobe and 
starts transferring its hydrogen envelope to the secondary, at which point
there are two possibilities, depending on the ratio of the thermal 
timescales of the two stars, which, in turn, depends on their 
mass ratio. The first possibilty arises if the mass 
ratio is not so extreme (quantitative arguments are given in the next 
section) that the above two timescales are not different by more than 
an order of magnitude, say. The secondary can then accept the entire 
mass transferred by the primary. This leaves behind only the He-core 
of the primary, and the secondary becomes so massive after
assimilating the H-envelope of the primary that it in fact turns into 
the more massive component of the system. This is the channel through
which HMXBs form. The second posibility arises if the mass ratio is 
so extreme that the thermal timescales are disparate by more than the
above amount, in which case the secondary is unable to accept the mass
transferred by the primary, and this mass forms a common enevelope (CE)  
surrounding the secondary and the He-core of the primary. This is the 
standard channel through which LMXBs are thought to form, which we 
do not discuss further in this paper.       

In the HMXB channel, the system now consists of a He-core plus a 
massive star, which was the secondary earlier, but is now the 
more massive component.
The He-core finishes the rest of its evolution rapidly 
(in $\sim 10^5-10^6$ years), and explodes in a supernova, leaving
behind a neutron star which will become the main X-ray emitter in 
the HMXB. The massive companion to the neutron star now finishes its
main-sequence evolution in $\sim 10^6$ years, expands to become a
supergiant, and drives a strong stellar wind. Accretion from this 
wind by the neutron star generates X-rays, and the system thus 
turns on as a HMXB. This HMXB phase lasts $\sim 10^4-10^5$ years,
after which the massive companion fills its Roche lobe. The 
subsequent mass-transfer is very large and rapid, which leads to 
considerable mass loss from the system, formation of a common
envelope (CE) surrounding the neutron star and the He-core of the
evolved massive companion, and a total extinction of the X-ray source.
The outcome of this CE evolution, \viz, the formation of a compact 
binary consisting of the neutron star and the He-core of its former 
massive companion, which, after the SN of this He-star, ultimately 
leads to the formation of a double-degenerate system consisting 
either of an eccentric double-neutron-star binary or of two runaway 
single neutron stars \citep{vdH01}, lies outside the scope of this
work.

In recent years, variations, modifications and additions to the 
above standard scenario have been considered. For example, it
has been suggested that CE evolution of sufficiently wide systems
during the first mass transfer may still lead to tight post-CE 
systems which would contribute to HMXB production (see \cite{linden} 
and references therein), and that these systems might remain 
tight even after the SN explosion, turning promptly into HMXBs with 
unevolved companions. Further, under certain other circumstances, CE 
evolution might lead to tightly-bound HMXBs with He-rich donors
\citep{linden}. In this first look at the problem, we shall not
consider these fascinating possibilities, but rather confine 
ourselves to the standard picture, since our aim here is to
assess the viability of our approach in the simplest testing 
ground, before attempting further refinements.
  
\subsection{Primordial Binary Distribution}
\label{sec:primordial}

We consider now the distribution of the essential properties of the 
primordial binaries from which HMXBs evolve according to the 
standard scenario sketched above. 
The progenitor primordial binary system is described by three
essential properties, namely, the mass of primary ($M_p$), the mass 
of the secondary ($M_s$), and the orbital separation ($a_0$). An 
equivalent description is in terms of $M_p$, the mass ratio 
$q = M_s/M_p$, and $a_0$. We use the second description throughout
this work, following the custom of recent literature on the subject,
and briefly mention the correspondence between the two descriptions
in Sec.\ref{sec:mdist}. Also, as is often done in this subject, we take the 
primordial binary orbits to be circular for the purposes of the
problem we study here. 

The constraints imposed by the HMXB problem under consideration here
limits the allowed range of the above parameters. If we restrict 
ourselves to only neutron-star HMXBs, as we do in this paper, then
the primary mass $M_p$ is restricted to be between 
roughly 9\Msun\ and 30\Msun\ \citep{Hegetal03}. 
This means, of course, that those HMXBs which have black holes (either
stellar-mass or intermediate-mass) as their compact, X-ray emitting 
components are outside the scope of this work. We return to this point
in Sec.\ref{sec:blackhole}. Next, the secondary mass $M_s$ is restricted 
from below by the requirement mentioned above, \ie, that its thermal 
timescale must not be larger than that of the primary 
by more than one order of magnitude, which
implies a rough lower bound $M_S\ge 0.3M_p$, since the thermal timescale
goes roughly as the inverse square of the mass in this mass-range. The
upper bound on $M_S$ is of course $M_s\le M_p$ by
definition. This means, therefore, that the allowed range of the 
primordial-binary mass ratio is roughly $0.3\le q\le 1$ for HMXB formation. 
Finally, the initial orbital separation $a_0$ is restricted 
from above by the condition of Roche lobe overflow when the primary 
evolves off the main sequence, without which there would be no mass
transfer, and the stipulated HMXB formation scenario will not operate.
It is customary to take this upper limit as $a_0 \le 10^3\Rsun$, 
which is not very restrictive.

Consider now the distributions of $M_p, q$, and $a_0$ we use in this
work, in keeping with the common practice in the literature.
The $M_P$-distribution is described in terms of a
suitable Initial Mass Function (IMF), \ie, the probability density
$f_M(M)$ of a stellar mass being in the range $M$ to $M+dM$. The IMF
is widely taken to be of a power-law form $f_M(M) \propto
M^{-\alpha}$, the classic Salpeter IMF \citep{Salp} corresponding to
$\alpha=2.35$, and a more recent suggestion being $\alpha=2.7$ 
\citep{Kroup}. We adopt the Salpeter IMF for our main calculations
and study the effect of varying the IMF on the
$L_X$-distribution in Sec.\ref{sec:IMFvary}. 

The $q$-distribution is generally taken as a power-law in $q$, 
given by $f(q)\propto q^{\beta}$, 
where the exponent $\beta$ may depend on the type of XRB
population being studied. Following the usage in recent works on 
population studies of HMXBs \citep{belc08,linden},  
we adopt a uniform $q$-distribution, $\beta = 0$, in our
HMXB studies in this paper. (However, we do note that our formalism 
has the provision for handling non-uniform distributions of $q$, which
will indeed be used in our LMXB studies described in subsequent 
papers in this series.) 

The distribution of the orbital separation 
$a_0$ is almost universally taken in the literature 
to be a loguniform one (\ie, an equal number of systems 
in equal intervals of $\log a_0$), also 
known as \"Opik's law \citep{Opik}. This implies a probabilty 
density $f(a_0)$ of the separation being 
in the range $a_0$ to $a_0+da_0$ as $f(a_0)
\propto 1/a_0$. We adopt \"Opik's law in our work. 

In order to construct the total probibility distribution function 
of primordial binaries, we need only note that the above pieces of
the probability density function (PDF) are independent of each other,
subject to the restrictions in the allowed ranges of values detailed
above. Thus, subject to these restrictions, the total PDF is 
given by:
\begin{equation}
f_{primo}(M_p, q, a_0) = \frac{1}{N} \frac{M_p^{-\alpha}}{a_0}
\label{eq:primopdf}
\end{equation}
Here $N$ is a normalization parameter. Note that, although this PDF 
is mathematically defined over much larger ranges of parameters, 
we are interested only in the allowed range for HMXB formation, as 
explained above. Hence, we set $f_{primo}=0$ over the forbidden range,
and so choose the normalization parameter $N$ that the integral of 
the PDF over the allowed range is unity. Consequently, although this
PDF has no explicit dependence on $q$ due to the assumed uniform 
$q$-distribution for this HMXB study, the allowed range of $q$, 
\ie, $0.3-1$, does have an effect on the value of $N$.

\section{Evolutionary Changes in Binary Parmeters}
\label{sec:paramchange}

In this section, we summarize the changes in the binary parameters as
the system evolves through the first mass-transfer phase, 
and the subsequent SN of the He-core of the original 
primary. As mentioned in Sec.\ref{sec:intro}, these changes only require
keeping track of the relations between the initial and final states in
each of the above two processes, which makes the ensuing
transformations relatively simple.   

\subsection{First mass transfer}
\label{sec:masstrans}

In keeping with previous work on the subject, we assume that there is
negligible mass loss from the system during this process, \ie, the 
mass transfer is \emph{conservative}, so that the H-envelope of the primary 
is entirely transferred to the secondary, leaving behind its He-core of 
mass $M_{p,c}$ given by:
\begin{equation}
M_{p,c} = M_0 M_p^{1/\xi},
\label{eq:coremass}
\end{equation}
where typical values of $M_0$ and $\xi$ used in the literature 
are 0.073\Msun and 0.704 respectively, for a metallicity of 
$z\approx 0.03$. This is a commonly-used analytic approximation 
to the results of numerical stellar-evolution calculations (Ghosh 2007
and references therein).

Because the mass transfer is conservative, the binary parmeters 
$\bar{M}_p$, $\bar{M}_s$, and $\bar{a}$ at the end of it are related 
to the primordial binary parameters as:
\begin{eqnarray}
\bar{M}_p &=& M_{p,c}~, \nonumber \\
\bar{M}_s &=& M_p + M_s - M_{p,c}~, 
\label{eq:pImtr}
\end{eqnarray}
and 
\begin{equation}
\bar{a} = a_0 \left[\frac{M_p M_s}{\bar{M}_p \bar{M}_s}\right]^2~.
\label{eq:pIachange}
\end{equation} 

\subsection{Supernova}
\label{sec:sn}

The post-mass-transfer binary is detached. The He-core described above
evolves further and explodes as a supernova (SN), which leaves behind a 
neutron star of a typical mass of 1.4 \Msun\ and blows off the rest 
of the He-star. This sudden mass loss from the system alters the orbital 
separation and makes the orbit eccentric. We assume in this work that
the neutron star has the above mass in all cases. Neglecting any 
effects of the expanding supernova ejecta on the secondary, \eg, 
ablation, the secondary remains completely unchanged in this process. 
Since we assume that the mass of the neutron star is always fixed to
1.4 \Msun, it can no longer be used formally as a free parameter in the
description of the post-SN system. Therefore, we choose the orbital 
eccentricity $e$ as the third parameter, so that the
post-SN system is described by the companion mass ($M_c$), the semimajor
axis ($a$) and the eccentricity ($e$).

Natal supernova kicks introduce additional changes in
the post-SN parameters. It is generally believed today that the SN 
explosion need not have exact spherical symmetry. Many suggestions
have been given for the physical origin of this asymmetry, \eg, density
inhomogeneity in the pre-collapse core, anisotropic neutrino emission, 
unequal momentum fluxes in the jet and anti-jet directions, and so on
\citep{scheck04, scheck06}. It has also been 
pointed out that a seed anisotropy, once introduced by these 
mechanisms, will be enhanced during the hydrodynamic evolution of the
explosion and may impart a large kick to the neutron star. 
Proper motion measurement of radio pulsars is used as a diagnostic of
these natal kicks. Such studies have shown that the distribution of
the pulsar velocities can be described by a 3D isotropic Maxwellian, 
in which the kick speed $v$ is distributed as
$f(v) \propto v^2 exp(-v^2/2\sigma^2)$ \citep{hobbs}, with
$\sigma = 265$ km/s \citep{hobbs}, and the direction of the kick 
velocity is distributed isotropically. Earlier works had suggested 
that the distribution might actually be bimodal, represented by two
Maxwellians, so that the overall distribution would be the weighted
sum of these \citep{arzoumanian}. 

Recent works have argued that electron-capture
supernovae (ECSN) are likely to produce much lower kicks with 
$\sigma \approx 50$ km/s or lower \citep{linden}, 
while the larger value given in the previous paragraph
is appropriate for iron-core-collapse supernovae (ICCSN). 
We note here that studies of the proper motions of 
pulsars involve single pulsars, some of which would have come from a
single star collapse. For the rest, which were obtained
by disruption of binary systems, there would clearly be an observational
bias towards higher kick velocities, since it is these systems
which became preferentially unbound to yield the single pulsars
whose proper motions are studied observationally. However,
such effects are difficult to account for at this stage of 
understanding of the problem, and it is customary in the literature
to apply the inferred distribution of kick velocities directly to
the X-ray binary progenitors, as we have done here. 

Effects of isotropic Maxwellian SN-kicks in progenitors of X-ray 
binaries were studied in a pioneering work by Kalogera (1996). 
For our study here, we have devised a method related to that 
described in the above work, but designed specifically for our
purposes here. The method works as follows. If we define our
co-ordinate system such that, just before the SN explosion, the line 
joining the two stars is along the x-axis and the neutron star is 
moving in the positive z-direction with a keplerian orbital 
velocity $v_{orb}$, then a kick of 
magnitude $v_k$ in the direction $\hat{n}(\theta, \phi)$ would 
produce a change in orbital parameters given by:

\begin{eqnarray}
\frac{M_i^t}{a_i} + v_k^2 + 2v_kcos\theta\sqrt{\frac{M_i^t}{a_i}}
&=& M_f^t\left(\frac{2}{a_i} - \frac{1}{a_f}\right) \nonumber \\
a_i^2 \left[v_k^2sin^2\theta sin^2\phi + \left(\sqrt{\frac{M_i^t}{a_i}} 
+ v_k cos\theta \right)^2\right] &=& M_f^ta_f\left(1-e^2\right) 
\label{eqn:postsn}
\end{eqnarray}

Here the angles $\theta$ and $\phi$ are defined in the usual way,
$M^t$ denotes the total mass of the system, and the subscripts $i$ 
and $f$ denote the initial and final values (\ie, \emph{pre-SN} and 
\emph{post-SN} values). All quantities here are in solar units, with 
$v$ in the units of $\sqrt{G\Msun/\Rsun}$. 

Due to the Maxwell-distributed random SN-kicks given to the neutron star,
the one-to-one correspondence between the pre-SN and post-SN binary 
parameteres, which would have existed in the absence of such kicks, 
is broken now. 
Instead, we need to calculate a suitable \emph{average} effect of 
SN-kicks on the transformation between the pre-SN and post-SN 
parameters, which takes into account the underlying distribution of
the kicks. To this end, we average eqs.\ref{eqn:postsn} over the 
distribution of $v_k$, upon which the linear terms in $v_k
cos\theta$ vanish due to isotropy, and the distribution-averaged 
transformation equations become: 
 
\begin{eqnarray}
\frac{M_i^t}{a_i} + \vsqav &=& M_f^t\left(\frac{2}{a_i} - 
\frac{1}{a_f}\right) \nonumber \\
a_i^2 \left(\frac{M_i^t}{a_i} + \frac{2}{3}\vsqav \right)
&=& M_f^ta_f(1-e^2) 
\label{eqn:postsnav}
\end{eqnarray} 

Here, \vsqav\ is $v^2$ averaged over the kick-distribution. For a
Maxwellian distribution, it is given by:
 
\begin{equation}
\left<v^2\right> = \frac{\int v^4 exp(-v^2/2\sigma^2)dv}
       {\int v^2 exp(-v^2/2\sigma^2)dv}
\label{eqn:vsqav}
\end{equation} 

Those properties of \vsqav\ for a Maxwellian which we need for our work
are detailed in Appendix A. We assume here a combination of two  
Maxwellians with values of $\sigma$ appropriate for ICCSN and ECSN, as
described above. The way for determining the appropriate proportions
of these two components is described below in Sec.\ref{invtrans}.
For each of the two components, the parameter transformations are
also given in Appendix A.  

\section{Post-SN probability density function}
\label{sec:psnpdf}

\subsection{Jacobian formalism}
\label{sec:jacob}

Probability theory provides us with a method of transforming the PDF of a
set of variable to the PDF of another set of variables, if we know
the relations between the these two sets of variables. Let us assume 
that a system is described by a set of variables denoted by a vector
$\bar{X}$. A given specific process transforms this set of 
parameters to another set of (the same number of) parameters denoted by 
$\bar{Y}$. Both the forward transformation $\bar{Y}(\bar{X})$ and the
inverse transformation $\bar{X}(\bar{Y})$ are mathematically defined. 
The initial PDF as a function of $\bar{X}$, $f_X(\bar{X})$ then 
transforms as follows:

\begin{equation}
f_Y(\bar{Y}) = f_X(\bar{X}(\bar{Y})) 
  \left| \frac{\partial \bar{X}}{\partial \bar{Y}} \right|
\label{eqn:jacob}
\end{equation}

Here, $\left| \frac{\partial \bar{X}}{\partial \bar{Y}} \right|$ is called
the Jacobian determinant of the inverse transformation. The above 
theorem can now be applied to our problem in order to transform the 
PDF of the primordial binaries to the post-SN PDF.

\subsection{Inverse transformation and post-SN PDF}
\label{invtrans}

In order to apply the above formalism to our problem, we
need to invert the above parameter transformation relations,
which is straightforward. The Jacobian of this 
transformation ($J_{mtr}$) can also be readily calculated. 
In the two regions described in Appendix A, these transformations 
can be obtained and the Jacobian ($J_{sn}$) can be calculated 
explicitly, as detailed in Appendix A.
  
Working in either region, the post-SN PDF is given by:

\begin{equation}
\bar{f}_{psn}(M_c, e, a) = f_{primo} J_{mtr} J_{sn}
\label{eqn:psnpdf}
\end{equation}

The PDF of the parameters
of the post-SN binary, $M_c,e,a$, no longer factorizes into 
individual PDFs for these three parameters, unlike the situation for
the PDF of primordial binaries. This is as expected, since 
the parameter transformation laws given above mix the 
parameters. An immediate consequence of this is that the PDF for any
one of the above post-SN binary parameters needs to be obtained by 
integrating the PDF of Eqn.\ref{eqn:psnpdf} over the other two
parameters. 

As described above, we use two values of the dispersion in the kick
distribution, $\sigma$, appropriate for ICCSN and ECSN. We  
have also described above the method of 
calculating the distribution in each case.
The actual distribution of the parameters with both types of SN present
is the weighted sum of these two PDFs for the two values of $\sigma$,
and this weighting comes from the following considerations.
For the range of $M_p$ which leads to ECSN events, we have adopted
here the value advocated in recent works \citep{pod04}, 
which is 8-11 \Msun .
The relative contribution of the ECSN and ICCSN is then 
given by $N_{EC}/N_{ICC} = N(M_{min},M_{tran})/N(M_{tran},M_{max})$, which 
depends upon the IMF. Here, $M_{min} = 9\Msun, M_{max} = 30\Msun$
define the allowed range of primary masses given earlier, and 
$M_{tran}$ is the primary mass for transition from ECSN to ICCSN.  
For a Salpeter IMF, this ratio is 0.42 for $M_{tran} = 11\Msun$ 
and 0.67 for $M_{tran} = 12\Msun$. We take this ratio as 0.42 
wherever it occurs in our calculations here.

\subsection{Properties of the post-SN PDF}
\label{sec:psnprop}

In order to appreciate the nature of the post-SN distribution, we 
calculate and plot the individual PDFs for each of these parameters,
obtained by integrating $\bar{f}_{psn}$ over the other two.
We now discuss each of these distributions.

\subsubsection{Companion mass}
\label{sec:compmass}

\begin{figure}
\includegraphics[scale=0.6,angle=270]{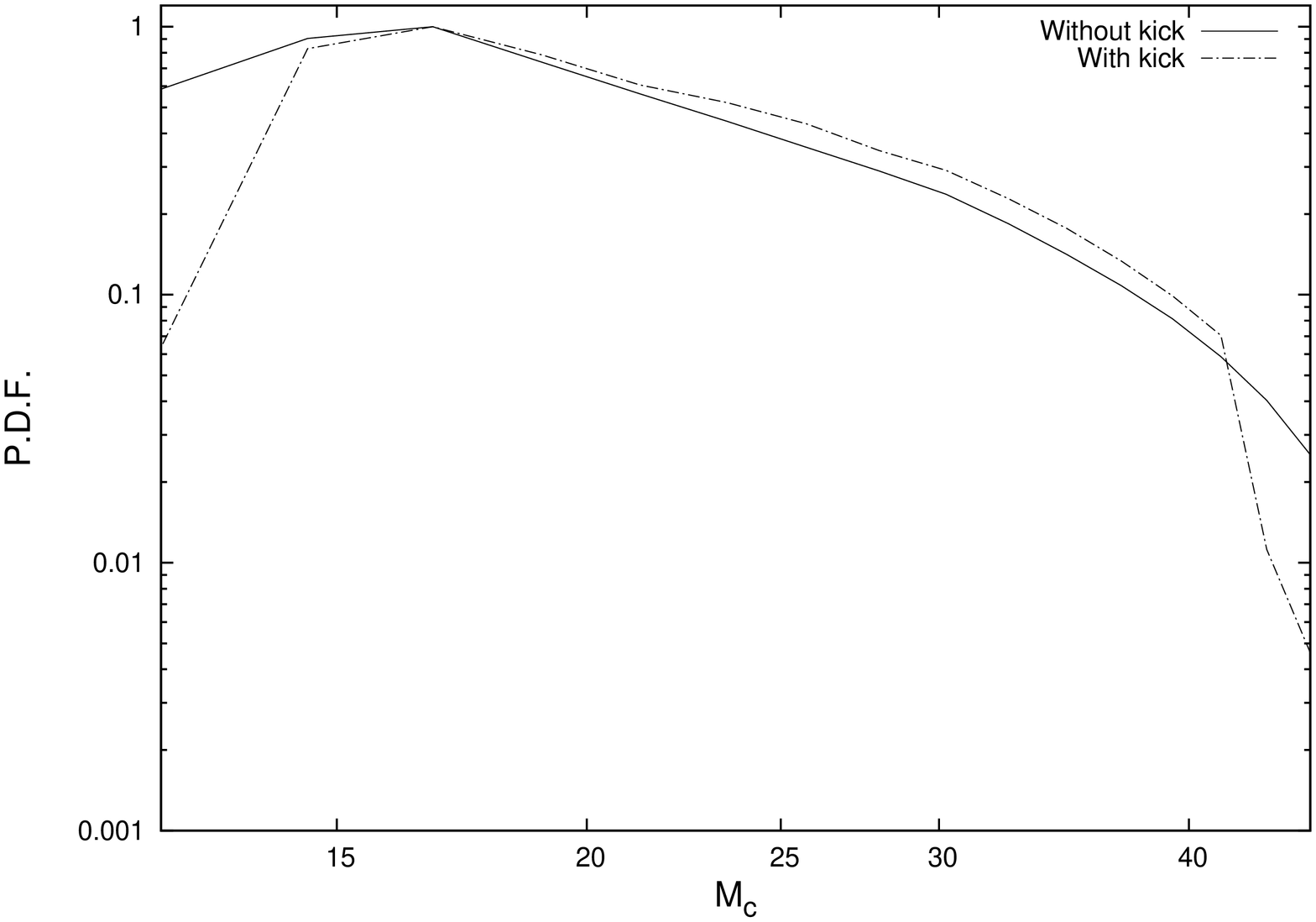}
\caption{Distribution of companion masses in post-SN systems.}
\label{fig:mcdf}
\end{figure}

\begin{figure}
\plottwo{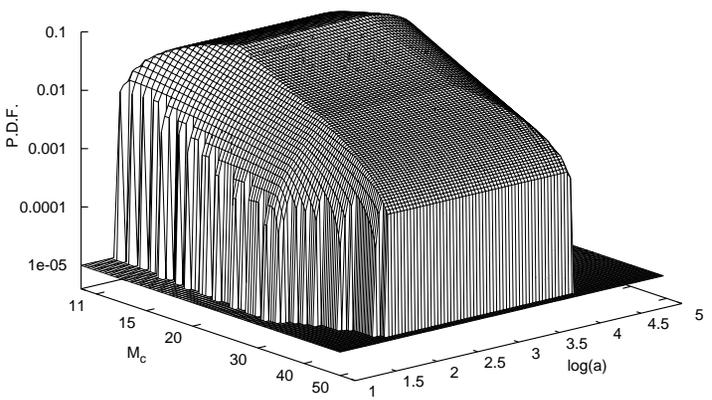}{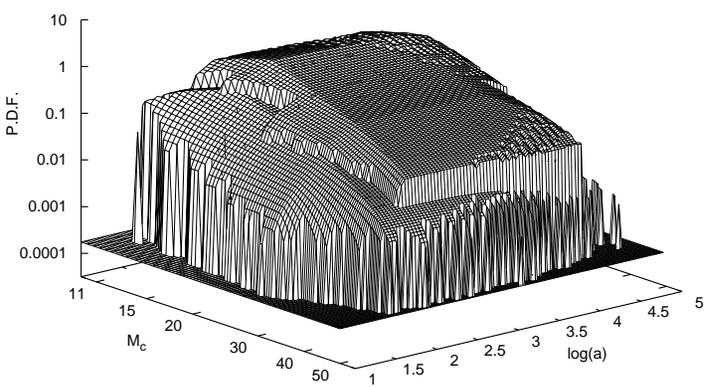}
\caption{Bivariate distribution of post-SN systems as a 
            function of both $M_c$ and $a$. Left panel: Without 
            SN-kicks. Right panel: With SN-kicks included.}
\label{fig:mcadf}
\end{figure}

The distribution of the companion mass shows a slow early rise and then
broken power-law behaviour with the break occuring at $\sim 30\Msun$, as 
shown in Fig.\ref{fig:mcdf}. The power-law slope is $\approx -2.4$,
in the mass range $20-30 \Msun$, whereas above the break point the
$M_c$-distribution becomes steeper, with a slope of 
$\approx -5.2$. The kink seen 
at the breakpoint of 30\Msun\ becomes a \emph{fold} in the
bivariate distribution as a function of $M_c$ and $a$ (See Fig. 
\ref{fig:mcadf}). This kink appears at the mass which
corresponds to the upper limit of the primary mass in the primordial 
binary. The allowed range of the primary mass is $9\Msun\leq M_p \leq 
30\Msun$: this constraint was imposed to ensure the formation of 
neutron star as an end product of the stellar evolution of the 
primary. 

It can be easily seen from the relation given above between 
the mass of the primary and that of its He-core that 
the mass range obtained for $M_c$ if we take $M_p = 30 \Msun$,
corresponding to the entire allowed range 
of $M_s$, is $\approx 30\Msun - 51\Msun$. Thus
for $M_c \leq 30 \Msun$, a large range of $M_p$ can contribute,
whereas for $M_c>30 \Msun$ the allowed range of $M_p$ is rapidly
cut off as $M_c$ increases, which causes a rapid reduction in the 
number of systems possible in this region. Thus, the effect of 
the allowed phase-space region 
is reflected in the transition in the power-law index of the 
$M_c$-distribution. In the region where large parts of the  
phase space of $M_p$ are allowed to contribute, the 
slope of the PDF is close to the input IMF slope of $-2.35$. 
For masses above the kink, only small parts can contribute, 
resulting in a steeper fall. For similar reasons of forbidden 
phase space, a downturn is expected at very low values of $M_c$,
in the mass range $\sim (10-15)\Msun$.

Upon inclusion of SN-kicks, the above mass distribution is 
essentially unchanged in the mid-$M_c$ range, except that 
the kink at $\sim 30\Msun$ becomes less prominent. But  
the distribution is cut off more sharply at the lowest and 
highest masses, the former being immediately understandable  
since kicks tend to preferentially unbind systems with 
lower binding energy. 
 
The bivariate $M_c-a$ distribution shown in Fig.\ref{fig:mcadf}
reveal the complexity introduced by the kicks. Although the 
surface shows similarities to its no-kick counterpart 
in general features, several additional features 
appear due to Maxwellian-averaged kicks corresponding to 
the two different values of $\sigma$ given
above for the two types of SN, namely, ECSN and ICCSN.

\subsubsection{Semimajor axis}
\label{sec:semimajor}

\begin{figure}
\includegraphics[scale=0.6,angle=270]{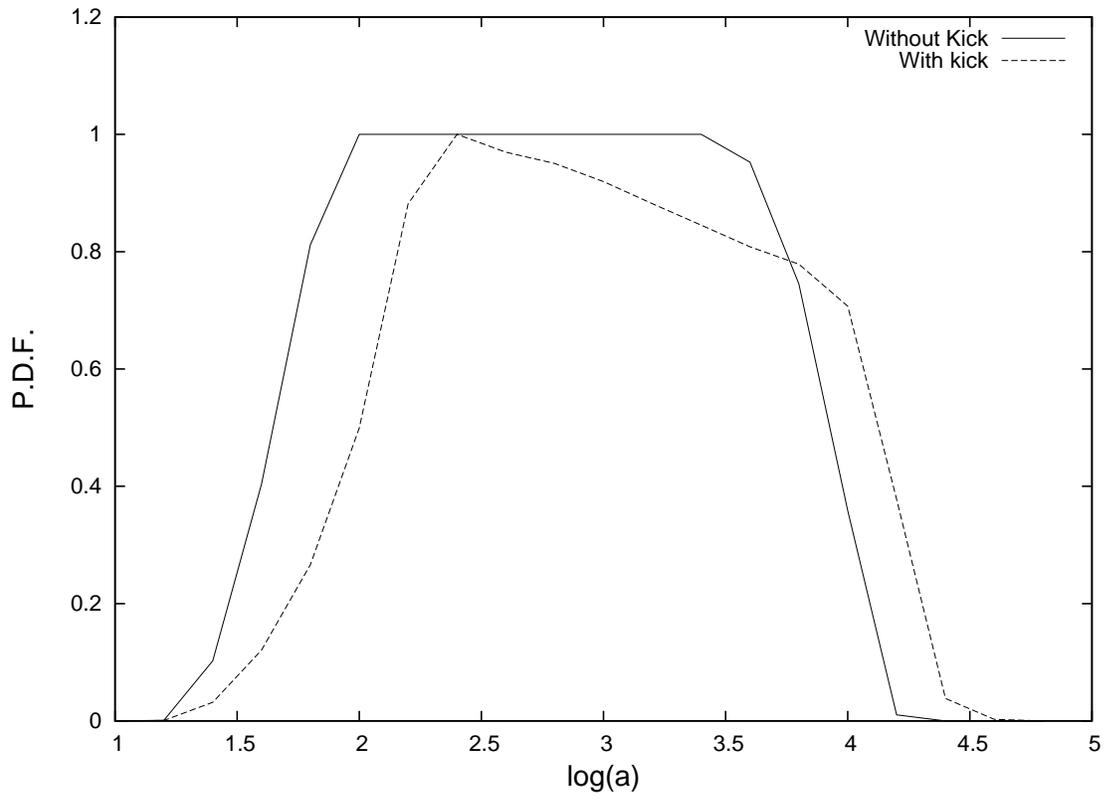}
\caption{Distribution of semi-major axes of post-SN systems.}
\label{fig:ladf}
\end{figure}

Fig.\ref{fig:ladf} shows the PDF of $log(a)$, the post-SN semimajor
axis on a logarithmic scale. The PDF without considering the effect of
SN-kicks is flat (\ie, loguniform in $a$) in the mid-$a$ range, 
with smooth rise and fall at 
low and high values of $a$ respectively. This is clearly a direct 
consequence of the assumed flat, \"Opik's-law distribution of  
primordial binaries, which remains unchanged
at the intermediate values of $a$. The PDF with the SN-kicks 
included clearly shows two additional features. First, there is
a general shift of the distribution towards wider orbits, entirely
as expected, since SN-kicks generally tend to widen orbits.  
Second, the PDF is not exactly loguniform, \ie, flat in the 
mid-$a$ range now: a slow fall with increasing $a$
is observed in this range, before the PDF falls off rapidly at 
large $a$, as before. This second feature is a consequence of the 
fact that SN-kicks steadily reduce the probability of survival of 
wider binaries even in the mid-$a$ range, before 
destroying them altogether at large $a$-values.

\subsubsection{Eccentricity}
\label{sec:eccen}

\begin{figure}
\includegraphics[scale=0.6,angle=270]{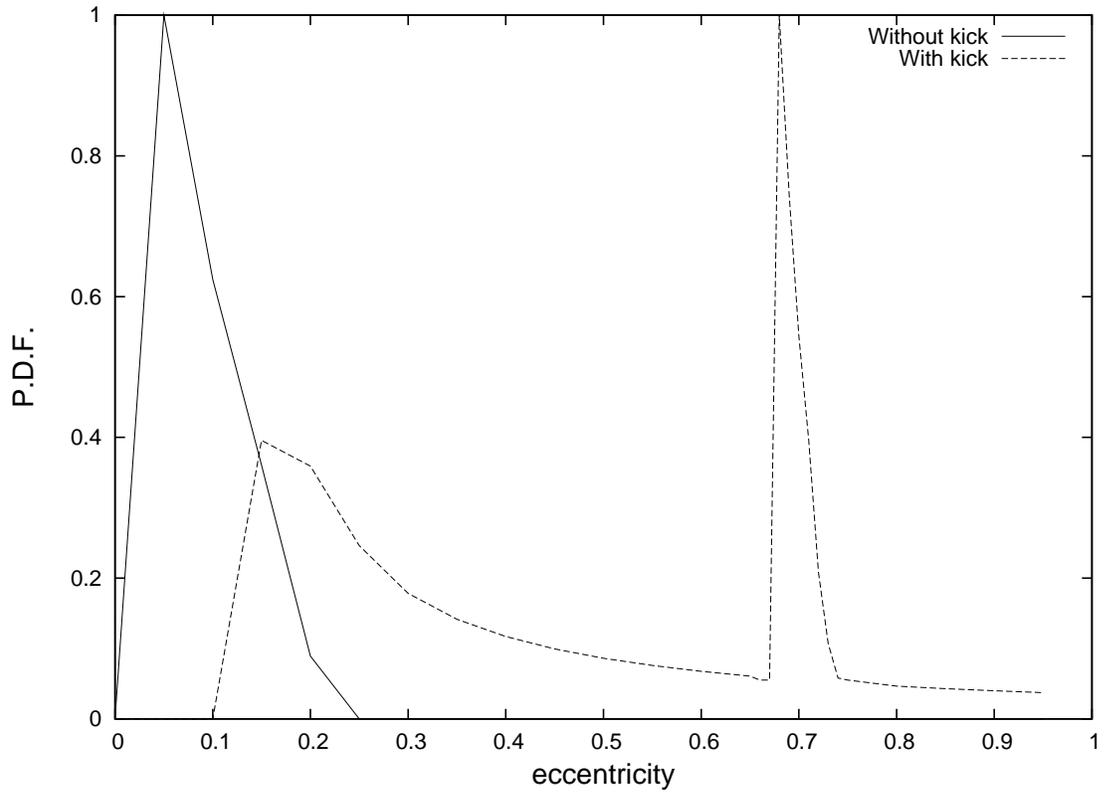}
\caption{Distribution of eccentricities of post-SN systems.}
\label{fig:edf}
\end{figure}

In the absence of kicks, the eccentricity introduced by the 
SN explosion is equal to the fractional mass loss form the 
system. As this fraction is small ($\sim 10$\% or less) in 
case of HMXBs, only small eccentricities are introduced by the 
SN in the no-kick scenario. Figure \ref{fig:edf},
which shows the post-SN eccentricity PDF, confirms this.
However, introduction of the SN-kicks
changes this result completely, making the post-SN systems
much more eccentric, as expected. Further, the results
for ICCSN and ECSN are quite different in terms of post-SN
eccentricity, also as expected because of the large difference
between $\sigma$ in the two cases (see above). Due to the small 
kicks imparted in case of ECSN with \vsqav\ independent of the 
details of the pre-SN system (see Sec.\ref{sec:sn} and Appendix A), the  
eccentricities are generally smaller than in the ICCSN case
(although larger than in the no-kick case), and the e-distribution 
is spread over a large range. By contrast, for ICCSN the kicks are
much larger, leading to large eccentricities, as shown in  
Fig.\ref{fig:edf}. Further, in this case \vsqav\ is almost a
fixed fraction of the disruption velocity of the pre-SN system 
for Maxwellian (and possibly also for other similar) distributions, 
as explained in Appendix A, so that the e-distribution 
is narrowly peaked around a value determined by that fraction.
It is likely that more detailed calculations would widen this peak
somewhat, but no qualitative changes are expected. 
The composite PDF of eccentricity for ECSN and ICCSN naturally
shows a double-peaked structure, as in Fig.\ref{fig:edf}.

Of course, the e-distribution calculated above refers to 
the immediate post-SN systems, which are not observable as 
XRBs. These systems subsequently undergo tidal 
circularization rapidly, so that most of them become circular
or nearly so by the time mass transfer begins and the 
system turns on as a HMXB. 
Since eccentricity is an irrelevant parameter at the HMXB stage,
we integrate $\bar{f}_{psn}$ over the eccentricity to obtain 
only the bivariate PDF as a function of $M_c$ and $a$, which is shown
in Fig.\ref{fig:mcadf}.

\section{Post-SN binary to HMXB}
\label{sec:psntoxrb}

Post-SN systems evolve over the main-sequence lives of their 
companions as detached systems on a short timescale
($\sim 10^6$ years), during which the orbit is tidally 
circularized. Therefore, a bivariate post-SN PDF adequately 
describes the HMXB system, being given by: 

\begin{equation}
f_{psn}(M_c, a) = \int_0^1 \bar{f}(M_c, a, e) de
\label{eqn:psnbi}
\end{equation}

Observed collective properties of HMXBs are usually given as 
distributions of their luminosities ($L$) (\ie, the X-ray luminosity 
function or XLF) and orbital periods ($P_b$), instead of $M_c$ and 
$a$, which we have worked with upto this point. In order to
compare our results with observations, a further transformation 
$(M_c, a) \rightarrow (L, P_b)$ is therefore required now.  
$P_b$ can be calculated for given values of $M_c$ and $a$ using Kepler's
third law. If all the masses and distances are expressed in the
units of the solar values, then $P_b$ in hours is given by:

\begin{equation}
P_b^2 = 7.72\, \frac{a^3}{M_c+M_{NS}}
\label{eqn:kepler}
\end{equation}

Here $M_{NS} = 1.4\Msun$ is the mass of the neutron star. To
calculate the luminosity as a function of $M_c$ and $a$, 
prescriptions of the mass-loss rate from the companion and the  
capture mechanism by the neutron star are required. 

The companion expands into a giant/supergiant after completing
its main-sequence life, and loses mass by driving a strong stellar 
wind at a rate $\dot{M}_w$. The neutron star captures a fraction 
of this lost mass. The accretion rate onto the neutron star 
is then given by $\dot{M} = \dot{M}_w \times$ (capture fraction). 
In this work, we assume that HMXBs are entirely 
\emph{wind-fed giant/supergiant} systems, thus ignoring those
systems in which a main-sequence companion close to its 
Roche lobe may drive a (weak) \emph{atmospheric} Roche-lobe
overflow. This is certainly justified at higher X-ray luminosities,
since systems fed only by the latter mechanism would be found at
the lowest end of the XLF. In a similar vein, we also do not 
include HMXBs with Be-star companions (see Sec.\ref{sec:bestar}), 
since the transient nature of these systems with low duty cycles
implies a low \emph{time-averaged} luminosity, which would put
them, again, at the low end of a time-averaged XLF, which is
generally difficult to determine.   
   
The X-ray luminosity $L$ is then given by the standard stellar-wind model,
according to which the massive companion drives a wind mass-loss at
a rate $\dot{M}_w$, and with a terminal wind speed of $V_w$. Of this,
a fraction $\sim\left[\frac{GM_{NS}}{V_w^2a}\right]^2$ (the capture
fraction introduced above) is captured and accreted by the neutron 
star, and the gravitational energy release from this accretion 
generates the X-rays. In order to connect $L$ 
to the stellar and orbital parameters, we first 
note that $L$ is the rate of release of the gravitational energy of 
the accreted matter and hence is directly proportional to accretion
rate \mdot as 
\begin{equation}
L = {GM_{NS}\mdot\over R_{NS}}~,
\label{eqn:acclum}
\end{equation} 
which we can express numerically as
$L_{36} \approx 1.17 \mdotten$, where $L_{36}$ is $L$  in
units of $10^{36}$ ergs/s and \mdotten is \mdot in units of 
$10^{-10}$ \Msun /yr. Next, we note that the accretion rate \mdot is
related to the wind mass-loss rate $\dot{M}_w$ and the above
capture fraction as 
\begin{equation}
\mdot = \dot{M}_w\times\left[\frac{GM_{NS}}{V_w^2a}\right]^2~.
\label{eqn:accrate}
\end{equation} 
To proceed further, we need models of stellar winds, which we 
consider next.   

\subsection{Stellar wind models}
\label{sec:windmodel}

Models of stellar wind relate the mass loss rate and the terminal
velocity of the wind to stellar parameters like mass, radius
and luminosity. In our case, these parameters are
the mass of the companion, the radius of the companion $R_c$
and the luminosity of the companion $L_c$. Some well-known 
stellar-wind models which we consider here are those due to 
Castor \etal, Kudritzi-Reimers, and Vink \etal\ 
\citep{castor, reimers, KudReim, vink}, which span the range from 
the classic CAK model \citep{castor} of the 1970s to the 
recent Vink \etal\ model \citep{vink}.

The rates of mass loss in these models are given by: 
\begin{eqnarray}
\mbox{Kudritzki-Reimers :} \;
\dot{M}_w &=& \gamma_{\rm KR} \: \frac{L_c R_c}
{M_c}\nonumber \\
\mbox{Castor \etal : } \;
\dot{M}_w &=& \gamma_{\rm CAK} \:  \frac{L _cR_c^{0.5}}
{M_c^{0.5}}  \nonumber \\
\mbox{Vink \etal} \;
\dot{M}_w &=& \gamma_{\rm Vink} \:  \frac{L _c^{2.45}R_c^{-0.5}}
{M_c^{1.3}}
\label{eqn:windmassloss}
\end{eqnarray}

The constants $\gamma_{\rm KR}$, $\gamma_{\rm CAK}$, and  
$\gamma_{\rm Vink}$ are given in the original works. We use 
the KR model in our main work here, and summarize the 
effects of varying the stellar-wind model in Sec.\ref{sec:windvary}.

\subsubsection{Stellar models}
\label{sec:starmodel}

Results of numerical calculations of stellar evolution
give stellar parameters like radius and luminosity as  
functions of the mass, metallicity and the age of the star.
For using the above stellar-wind models in our calculations,
we need these parameters for a given mass and metallicity of
the wind-driving companion. In the spirit of our semi-analytic
approach, we use the parameters given by Hurley, Pols and 
Tout (2000, henceforth HPT) in their comprehensive work on
the construction of analytic approximations to the standard
numerical results of stellar-evolutionary codes. The range 
of $M_c$ shown in Fig.\ref{fig:mcdf} and discussed in 
Sec.\ref{sec:compmass} leads to the conclusion that, except 
for a small region at the lower end of this range, the massive 
companions ignite He in the core while in the Hertzsprung  
gap (HG), while those in that small region do so at the top 
of the giant branch (GB). The companion's luminosity remains 
nearly constant in the HG, as discussed in HPT. Relevant
fitting formulae for the luminosity and radius in the 
core Helium burning (CHeB) phase are given in Sec.5 of HPT, 
particularly Secs.5.1 and 5.3. We have used a metallicity 
$z=0.02$ in our main calculations, varying it later to test its  
effect on HMXB distribution (see Sec.\ref{sec:othervary}).

The terminal velocity $V_w$ which appears in Eqn.\ref{eqn:accrate}
is of the order of escape velocity from the surface of the 
companion and hence can be taken as a function of $M_c$, so
that the accretion rate can be expressed in the form:

\begin{equation}
\dot{M} \propto \frac{f(M_c)}{a^2},
\label{eqn:accr}
\end{equation}

where the function $f(M_c)\sim\dot{M}_w/V_w^4$ contains
the physics of wind mass-loss. Equation \ref{eqn:accr}
gives the accretion rate as a function of $M_c$ and $a$ and
so, along with Eqn.\ref{eqn:kepler}, provides the transformation 
from $(M_c, a)$ to $(\dot{M}, P_b)$, 

\subsection{Transformation of parameters}
\label{sec:hparamtr}

Inverse transformations for eqns. \ref{eqn:accr} and 
\ref{eqn:kepler} need to be obtained to transform the PDF. We
first carry out the transformation from $(M_c, a)$ to 
$(\dot{M}, P_b)$. We note here that the PDF as a function
of $\dot{M}$ and $L$ will be identical up to a numerical factor since
the two are linearly related to each other. The inverse transformation 
can become complicated because $f(M_c)$ can be a very complicated 
function and hence difficult to invert analytically. Our goal would,
of course, be to construct a procedure that works with a general 
$f(M_c)$, so that the formalism can be applied to any model without 
changing the procedure. We first write the accretion rate as

\begin{equation}
\mdotten = 3.92 \frac{f(M_c)}{a^2} .
\label{eqn:acrate}
\end{equation}

The numerical factor is so chosen that $\gamma_{\rm KR}=10^{-4}$;
for other models the numerical constants can be appropriately 
adjusted in a similar way. With the aid of Eqns.\ref{eqn:kepler} and 
\ref{eqn:accrate}, we can eliminate $a$ and write

\begin{equation}
g(M_c) = G(P_b, \mdotten) ,
\label{eqn:htr1}
\end{equation}

where

\begin{eqnarray}
g(M_c) &=& \frac{f^3(M_c)}{(M_c+M_{NS})^2} .\nonumber \\
G(P_b, \mdotten) &=& 2.78 \times 10^{-4} \mdotten^3 P_b^4 .
\label{eqn:defhtr1}
\end{eqnarray}

This equation can be numerically solved to obtain $M_c$ as a
function of $(P_b, \mdotten)$, which is the first inverse 
transformation equation. After computing $M_c$ numerically, 
$a$ can be computed using

\begin{equation}
a = h(M_c) H(P_b, \mdotten),
\label{eqn:htr2}
\end{equation}

where

\begin{eqnarray}
h(M_c) &=& \frac{M_c+M_{NS}}{f(M_c)} .\nonumber \\
H(P_b, \mdotten) &=& 3.3 \times 10^{-2} \mdotten P_b^2 .
\label{eqn:defhtr2}
\end{eqnarray}

Using these inverse transformations, we calculate the Jacobian. The partial 
derivatives take the following forms in terms of the functions
$G(\dot{M}_{-10}, P_b)$, $H(\dot{M}_{-10}, P_b)$, $g(M_c)$ and $h(M_c)$:
\begin{eqnarray}
\frac{\partial M_c}{\partial \mdotten} &=& 
\frac{\partial G}{\partial \mdotten} \left(\frac{dg}{dM_c}\right)^{-1}
\nonumber \\
\frac{\partial M_c}{\partial P_b} &=&
\frac{\partial G}{\partial P_b} \left(\frac{dg}{dM_c}\right)^{-1}
\nonumber \\
\frac{\partial a}{\partial \mdotten} &=& h\:\frac{\partial H}{\partial \mdotten}
 + H\:\frac{dh}{dM_c}\left(\frac{dg}{dM_c}\right)^{-1}\frac{\partial G}{\partial \mdotten}
\nonumber \\
\frac{\partial a}{\partial P_b} &=& h\:\frac{\partial H}{\partial P_b}
 + H\:\frac{dh}{dM_c}\left(\frac{dg}{dM_c}\right)^{-1}\frac{\partial G}{\partial P_b}
\label{eqn:partials}
\end{eqnarray}

From eqs.(\ref{eqn:htr1}), (\ref{eqn:htr2}) \& (\ref{eqn:partials}), we can 
calculate the Jacobian as a function of $P_b$ and \mdotten\ as:

\begin{equation}
J = \left[\frac{M_c+M_{NS}}{f(M_c)}\right]^3\;
\left\{3\:\frac{df}{dM_c}-2\:\frac{f(M_c)}{M_c+M_{NS}}\right\}^{-1}\;
\left[1.84\times10^{-5}\:\mdotten^3P_b^5\right]
\label{eqn:hjacob}
\end{equation}

\subsection{Transformation of Distributions}
\label{sec:transdist}

The distribution of HMXBs in $P_b$ and \mdotten\ can now be obtained 
with the aid of Eqs.(\ref{eqn:jacob}) and (\ref{eqn:psnbi}). It is:

\begin{equation}
f_{HMXB}(P_b, \mdotten) = f_{PSN}\left(M_c(P_b, \mdotten), 
a(P_b, \mdotten)\right)\;\left|J(P_b, \mdotten)\right|,
\label{eqn:hmxbpdf}
\end{equation} 

and $\mdot$ and $L$ are of course proportional to each other,
as explained above.

\begin{figure}
\plottwo{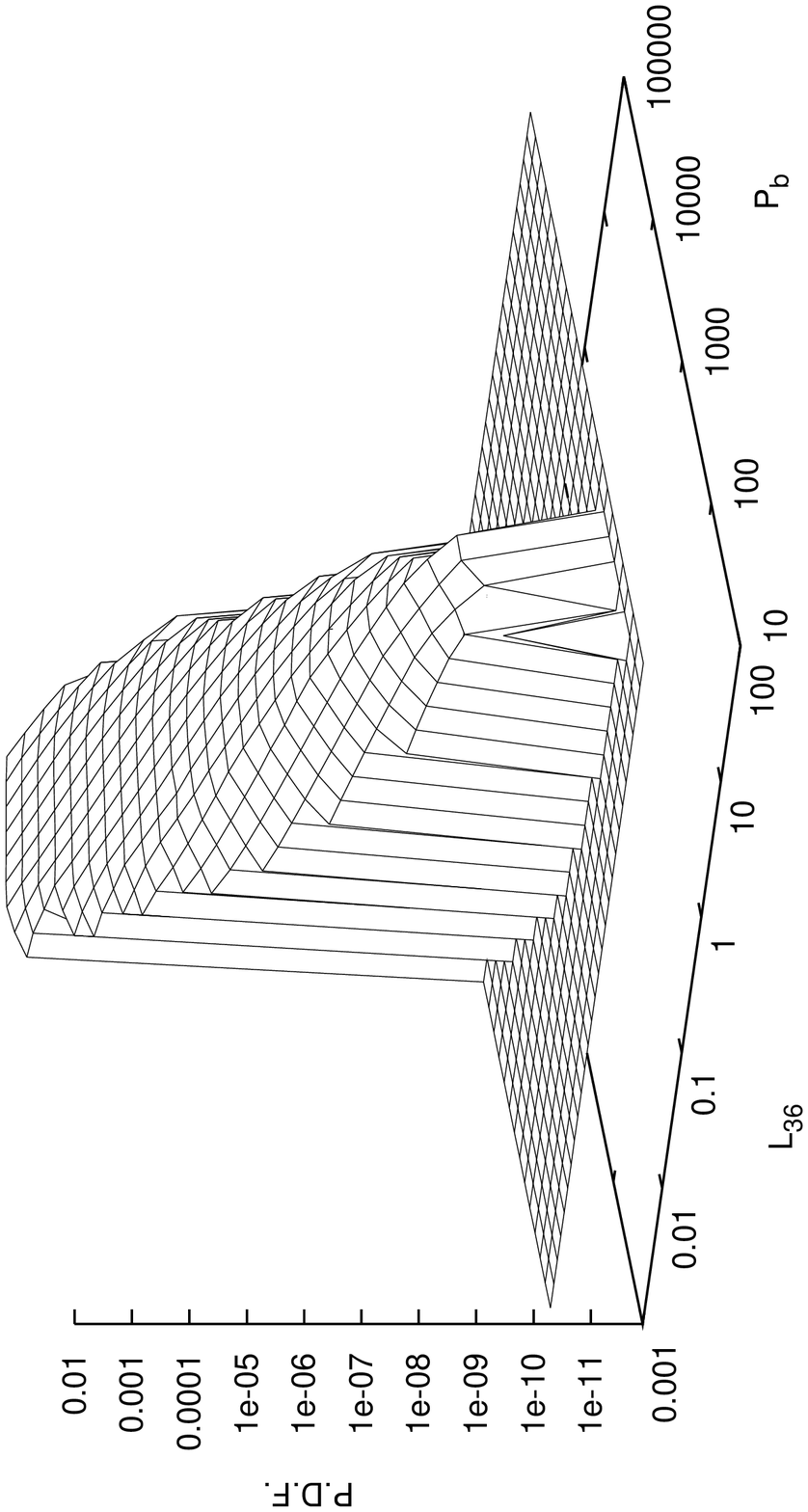}{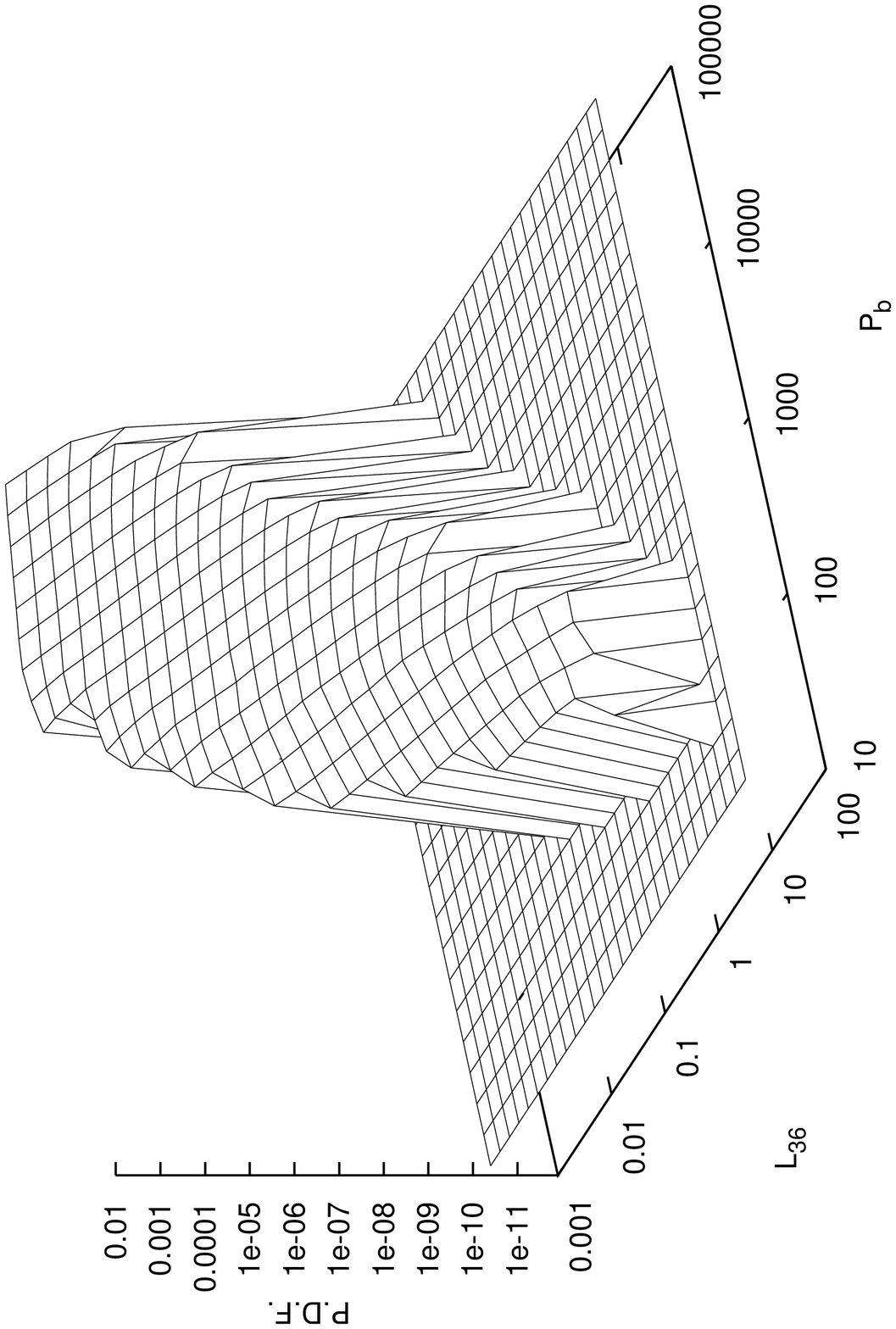}
\caption{Bivariate distribution of HMXBs as a 
            function of both $L$ and $P_b$, including the 
            effects of SN-kicks. Two views of the 
            3-dimensional figure are shown in order to bring
            out the trends with $L$ (left panel) and $P_b$
            (right panel).}
\label{fig:pbldf}
\end{figure}

\begin{figure}
\includegraphics[scale=0.6,angle=270]{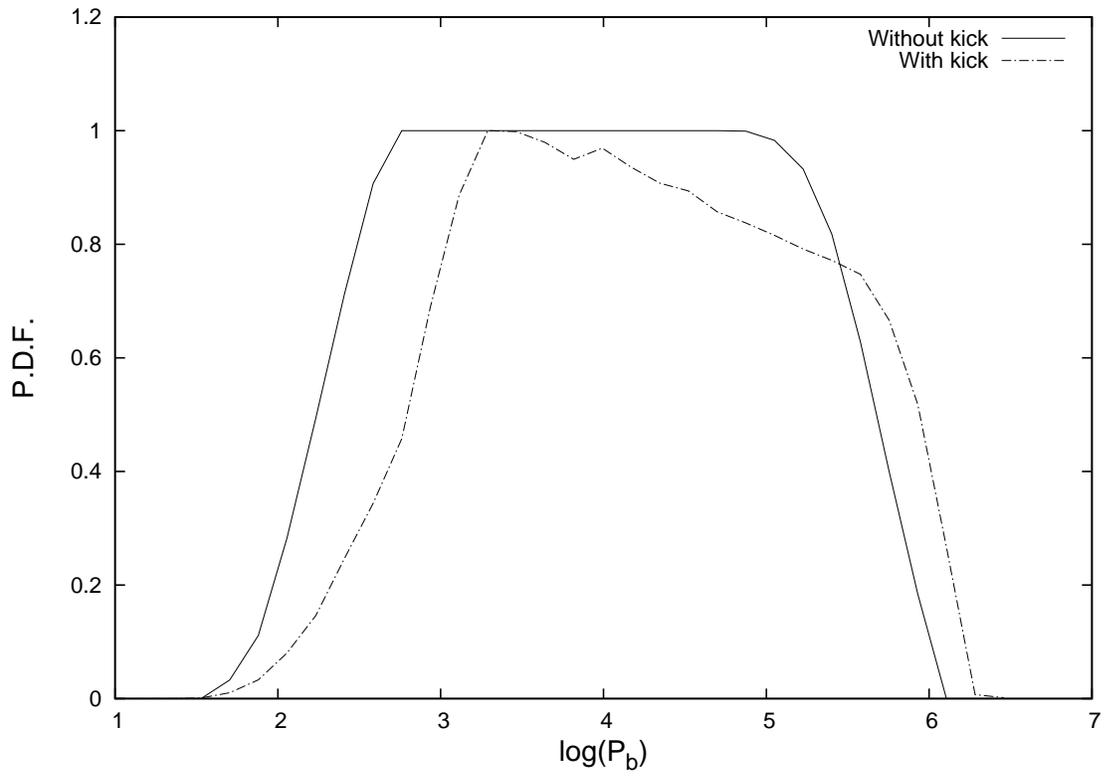}
\caption{Calculated distribution of HMXB orbital periods.}
\label{fig:pdist}
\end{figure}

The resultant bivariate PDF  
$f_{HMXB}(L,P_b)$ is displayed as a surface in Fig.\ref{fig:pbldf}, 
and the individual PDFs for the $L$- and $P_b$-distributions are shown 
in Figs.\ref{fig:hxlf} and \ref{fig:pdist} respectively. 
For obtaining the individual 
distribution of each variable, the bivariate distribution is
integrated over the other variable, as explained earlier, the range
of integration extending over only the allowed range of the concerned
variable.

\begin{figure}
\includegraphics[scale=0.6,angle=270]{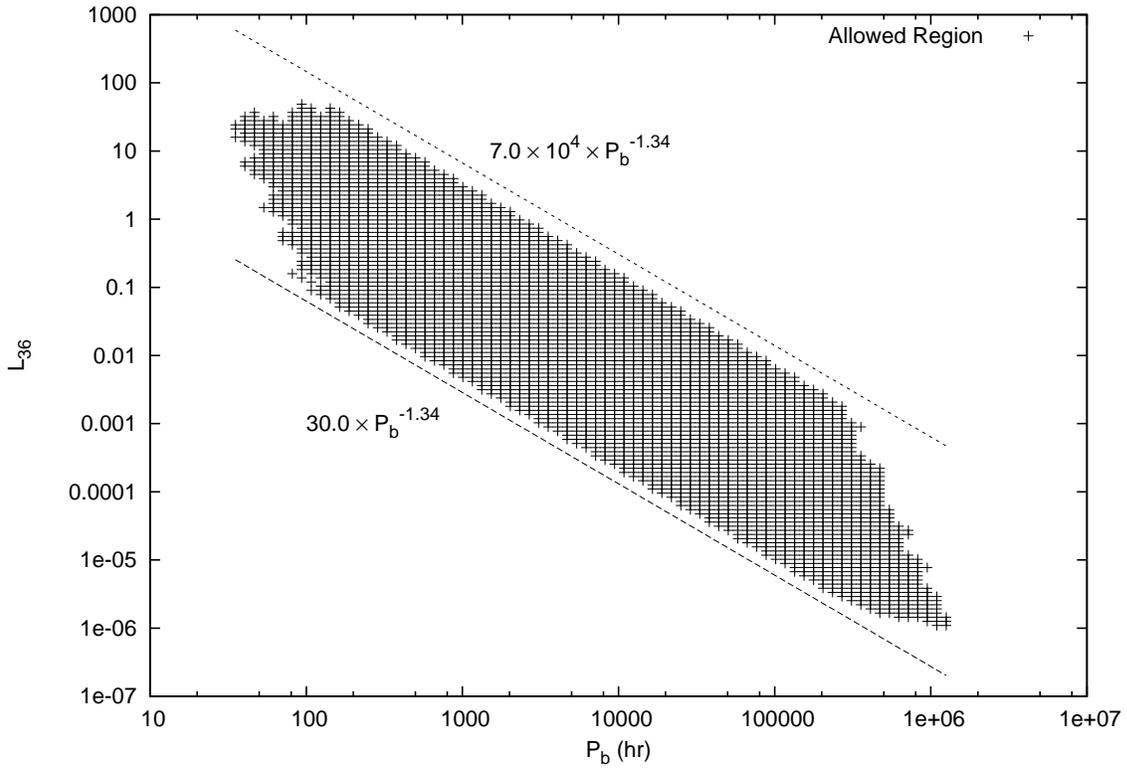}
\caption{Allowed zone in $L-P_b$ space shown covered by 
    crosses. Upper and lower boundaries of zone as indicated.}
\label{fig:pblrange}
\end{figure}

In Fig.\ref{fig:pblrange}, we show the allowed region in the $L-P_b$ 
plane, which shows how the allowed region in the $M_c-a$ plane shown in 
Fig.\ref{fig:mcarange} is mapped onto the plane of the new variables. 
The effect of the above transformation is a
rotation of the allowed zone in $(P_b, L)$ plane. The simple bounds
on parameters of primordial binaries given earlier transformed into 
a nearly rectangular allowed region in the parameter space $(M_c, a)$
for post-SN binaries (see Fig.\ref{fig:mcarange}).
After transformation to the \emph{HMXB parameters}, however, this 
allowed zone rotates into an inclined band, as shown in
Fig.\ref{fig:pblrange}. The boundaries given in this figure serve 
as integration limits for the calculation of the individual PDFs 
from the bivariate distribution in terms of $L$ and $P_b$.

\section{The HMXB Distribution}
\label{sec:hmxbdist}

We now discuss the nature of the distributions of the luminosities and 
binary periods of HMXBs calculated above, and compare them with 
the current state of observational knowledge of these distributions.
We emphasize that we are not attempting a detailed fit to the data
at this stage, but focusing instead on a comparison between the
general trends in calculated and observed distributions. The purpose
of such a comparison is of course to  determine if a more detailed 
computational scheme, \eg, a numerical population synthesis, is worthwhile in 
the future for a giving a more detailed account of the observations.
A major virtue of an approach like ours is that it is 
capable, at least in principle, of assessing  in a transparent way the 
relative importance of various components (\eg, the role of the 
primordial binary distributions vis-a-vis that of the evolutionary 
processes like the first mass transfer and the SN) in determining 
the final HMXB distribution. Such an assessment helps greatly in 
planning the strategy of subsequent computational studies. 

\subsection{Orbital Period Distribution}
\label{sec:porbdist}

The PDF of HMXBs as a function of $P_b$ is obtained immediately
by integrating $f_{HMXB}$ given in Eqn.\ref{eqn:hmxbpdf} over the 
accretion rate. The computed $P_b$ distribution 
without SN-kicks (see fig. \ref{fig:pdist})
is rather similar to the corresponding post-SN $a$ distribution, 
\ie, a flat \"Opik's law in the mid-region ($P_b \sim 300 - 3\times 10^4$
hrs), with a gradual rise to this flatness at shorter periods and a 
slightly sharper fall-off from it at longer periods. 
The distribution with the inclusion of SN-kicks also resembles the post-SN
$a$ distribution with SN-kicks: the rise now is more gradual, reaching
a maximum at $P_b \sim 1000-1500$ hrs. Instead of a flat top, 
a slow fall is observed mid-$P_b$ range, 
followed by a sharper fall-off at long periods, as before. The major
and obvious difference between the two distributions is an overall 
shift towards wider orbits and longer periods when SN-kicks are
included, entirely as expected and as seen earlier in the post-SN
$a$-distribution. We compare our theoretically obtained
$P_b$-distribution with observations in Sec.\ref{sec:observe}.

\subsection{X-ray luminosity function}
\label{sec:hxlf}

The XLF of HMXBs can be obtained by integrating Eqn.\ref{eqn:hmxbpdf} 
over $P_b$, using the limits described previously and carrying out a 
simple linear transformation for \mdotten\ to \lum. The XLF 
is given by:

\begin{equation}
f_L(\lum) = \frac{1}{1.17} \int f_{HMXB}(P_b, \mdotten) \, dP_b
\label{eqn:hxlfgen}
\end{equation}

where $\mdotten = \lum/1.17$. Figure \ref{fig:hxlf} shows the numerically
computed XLF. A broken power-law can be fitted to this computed XLF, 
with a cut-off at the neutron-star Eddington limit. 
Equation \ref{eqn:hxlf} gives the 
power-law exponents in the low and high luminosity regimes, obtained
by least-squares fits to the computed XLF: 

\begin{equation}
\frac{dN}{dL}\propto\:L^{\eta}\; \mbox{with}\; 
\eta = \left\{\begin{array}{ll}
-1.63 & 0.65 < L_{36}< 7.5 \\
-0.98 & 10^{-3}< L_{36}< 10^{-2}
\end{array} \right.
\label{eqn:hxlf}
\end{equation}

The XLF calculated without SN-kicks is also given for reference. It 
overlaps with the XLF with SN-kicks in the low-luminosity regime, and 
shows a slightly shallower power-law of exponent is $\eta = -1.43$ in the
high-luminosity regime, the cross-over point being at
$L_{cr} \approx 8\times 10^{34}$ erg/s.

We now examine the effects of the various model parameters, \eg,
stellar-wind model, IMF slope, metallicity, and so on on our theoretical
XLF. 

\subsection{Comparison with observations}
\label{sec:observe}

\subsubsection{The XLF}
\label{sec:compxlf}

\begin{figure}
\includegraphics[scale=0.6,angle=270]{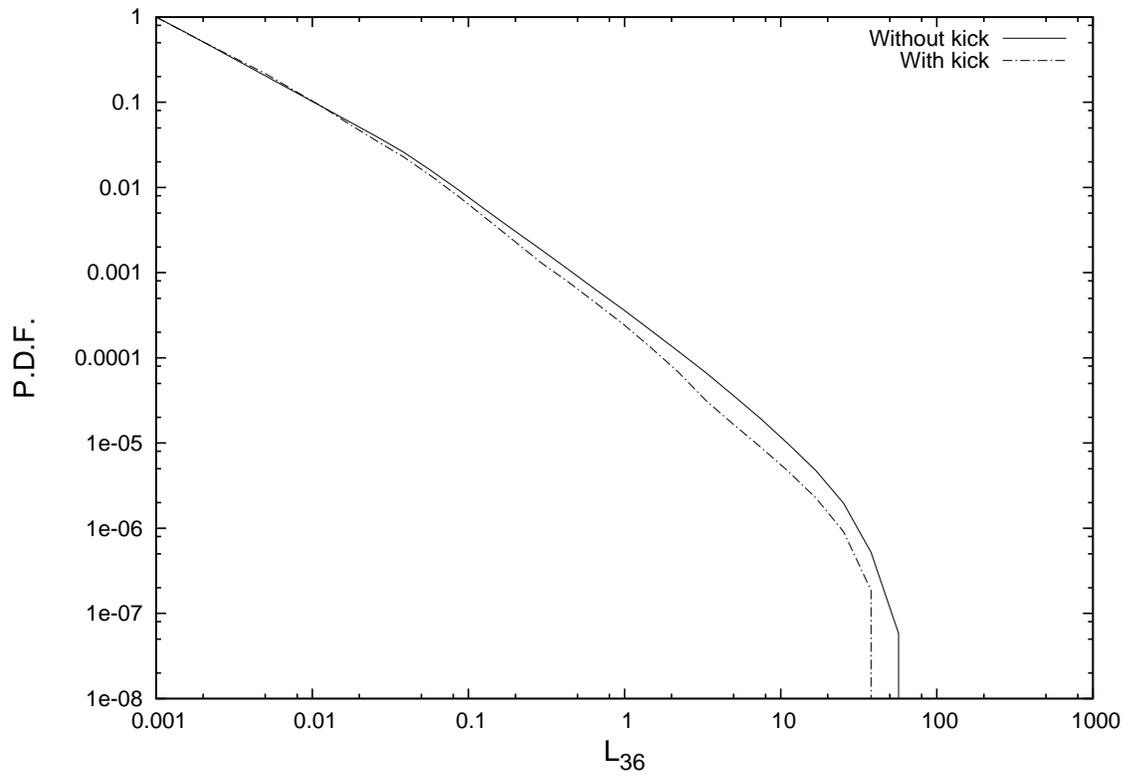}
\caption{Calculated XLF of HMXBs.}
\label{fig:hxlf}
\end{figure}

We first compare the calculated $L$-distribution of Fig. 
\ref{fig:hxlf} with the 
observed distribution. The major feature of the observed  
$L$-distribution is a power law 
with a differential slope $dN/dL\propto L^{-1.6}$ over a wide 
range of luminosities $3\times 10^{35}{\rm erg/s}
\le L \le 3\times 10^{40}{\rm erg/s}$. This is the so-called
``universal'' X-ray luminosity function (XLF) of HMXBs, obtained in
the following way. Whereas the observed XLFs of various nearby 
early-type galaxies (\eg, the Milky Way, SMC, M82 and M83, the Antennae, 
NGC 4736, and so on) follow this trend, their normalizations are 
not the same. However, when the XLF of a given galaxy is normalized 
by the current star-formation rate (SFR) in that galaxy, XLFs for all
these galaxies fall essentially on top of each other, yielding the
above ``universal'' XLF \citep{Gil04,GGS04a,GGS04b,Grimetal02,
Grimetal03}. This is due to the well-known fact that the 
strength of the HMXB population and their X-ray output is proportional
to the current SFR, and is closely related to the discussion given
in earlier on how the rapid HMXB evolution leads
to a distribution of HMXBs which is, in effect, a ``snapshot'' of 
the galaxy taken at its current SFR. 

In comparing calculated and observed XLFs, we note first that,
since we confine ourselves to only neutron-star HMXBs in this 
work, our calculated XLF applies only to luminosities not exceeding 
the Eddington luminosity $L_E$ for a $\sim$ 1.4\Msun\ neutron star, \ie, 
$\sim 2\times 10^{38}{\rm erg/s}$. The observed XLF extends upto 
almost 2 decades of luminosity above this, and the HMXBs at these higher
luminosities are believed to be black-hole systems, with the possibility
of both stellar-mass black holes and intermediate-mass black holes (at
the highest luminosities) being present. We return to this question in
more detail in Sec.\ref{sec:blackhole}.

With this caveat in mind, we note that there is a remarkable agreement 
between the calculated and observed XLFs in the luminosity range 
$3\times 10^{35}{\rm erg/s}\le L\le 10^{38}{\rm erg/s}$,
above which the calculated XLF cuts off at the Eddington limit
for neutron stars, as expected, so that no comparison with
observations is possible at higher $L$. Below $L\approx 3\times 
10^{35}{\rm erg/s}$, there was no data
when the works referred to at the beginning of this subsection were 
published. However, more recent $XMM-Newton$ observations of
the Magellanic Clouds have extended the XLF below this lower limit,
down to about  $L\approx 10^{34}{\rm erg/s}$ or slightly lower, for 
SMC and LMC \citep{SG05a,SG05b}. The results indeed suggest a flattening of  
the XLF at low luminosities, and the best-fit differential slope of
the observed XLF at these luminosities for SMC, which is given in 
the above reference
as $-1.13^{+0.3}_{-0.13}$, is in fact almost consistent with our 
calculated slope at low luminosities, given in Fig.\ref{fig:hxlf} and 
Eq.(\ref{eqn:hxlf}). However, we must be cautious with our calculated
XLF in the low-luminosity regime, since we have neglected (a)
main-sequence companions undergoing atmospheric Roche-lobe overflow, 
and (b) Be-star companions in our calculations here, as explained
above.

\subsubsection{The $P_b$-distribution}
\label{sec:comppb}   

\begin{figure}
\includegraphics[scale=0.6,angle=270]{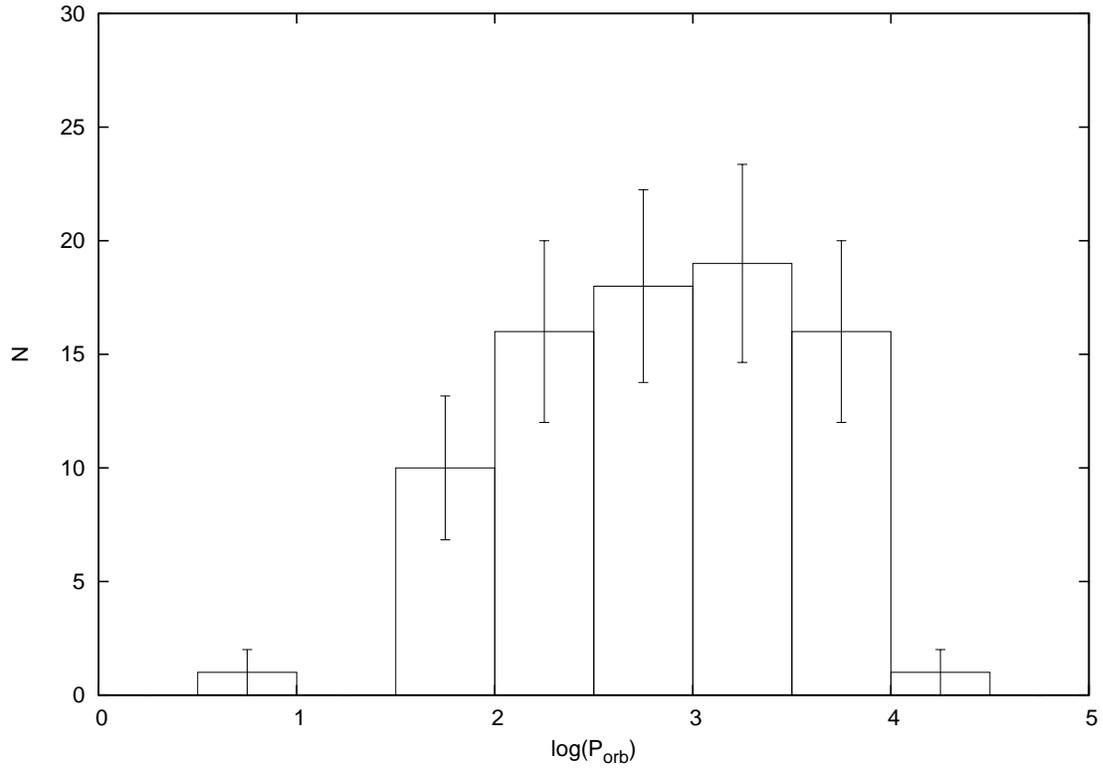}
\caption{Observed distribution of HMXB orbital periods.
               Shown is the data for $\sim 80$ systems in the  
               Milky way, SMC and LMC. From the catalogues by 
               Liu \etal: see text.}
\label{fig:pbobs}
\end{figure}

Observations of HMXB orbital periods are available in substantial 
numbers only for our galaxy and the Magellanic Clouds.  Measurements 
of orbital periods have been performed for $\sim 80$ systems 
\citep{LPH05, LPH06}.
Figure\ref{fig:pbobs} shows the distribution of orbital periods 
constructed with this data. The distribution shows a rising part in the range
$10-300$ hrs. With the (large) error bars, the distribution in the region 
$P_b \approx 300 - 10000$ hrs is consistent with either a uniform
trend or a slow rise/decay with a peak around $P_b \approx 2000$ hrs.  
It can be seen that our theoretical $P_b$-distribution is generally
consistent with the observed one within the error bars, except at very 
long orbital periods, where the observed distribution shows a sharp 
cut-off beyond orbital periods $\sim 1$ year.
Such an apparent cut-off is expected for two observational reasons. 
First, it is very difficult to follow systems with such long orbital
periods for sufficient times to establish reliable orbital periods. 
Second, such wide binary systems would generally have low
luminosities, which would put them at the faint end of the
XLF, and add further to the difficulties of a successful
observation. Thus strong selection effects work against observation
of HMXBs with long orbital periods. By contrast, the theoretical
distribution extends upto the widest orbits that can survive the 
basic processes involved, particularly the SN explosion. The fact
that these very wide binaries do not show up in the observed 
distribution does not imply that they do not exist, but simply that
they are unobservable in practice.

\subsection{Parameter Study}
\label{sec:parstud}

We now examine the effects of the various model parameters, \eg,
stellar-wind model, IMF, metallicity, and so on on our theoretical
XLF. 
 
\subsubsection{Effects of stellar wind models}
\label{sec:windvary}

\begin{figure}
\plottwo{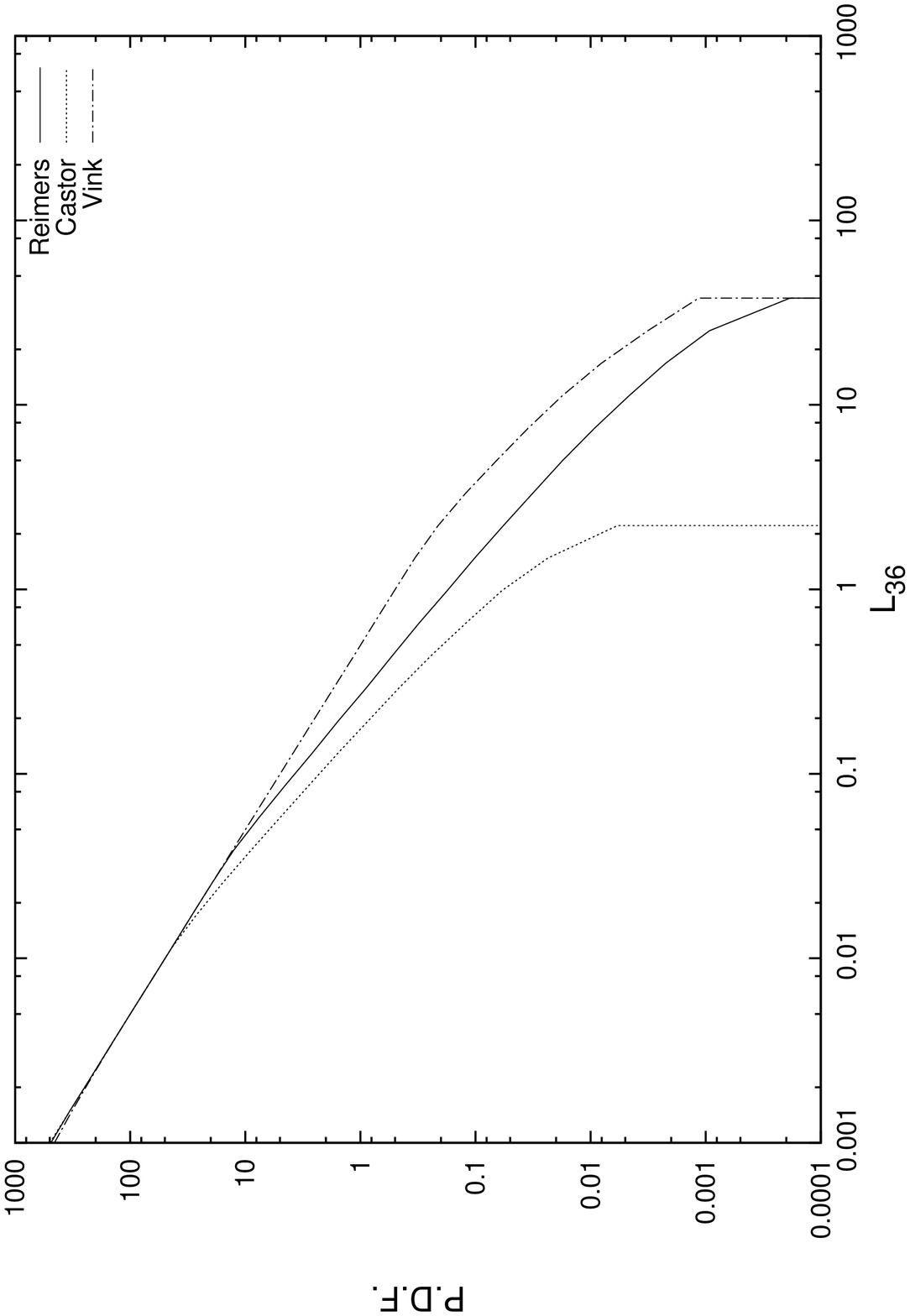}{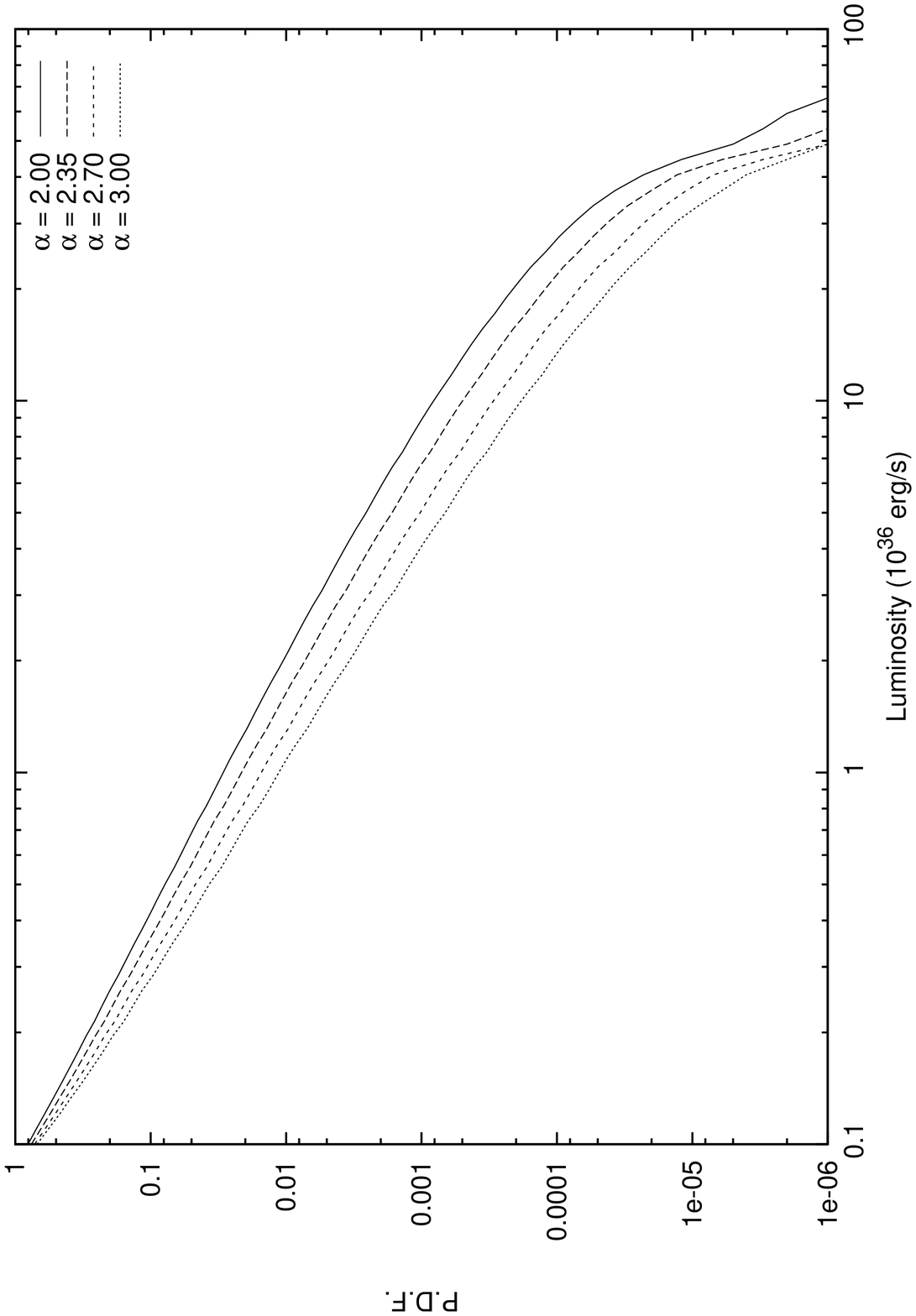}
\caption{Dependence of XLF on model parameters. Left panel:
                different stellar wind models, as labeled. 
                Right panel: different IMF slopes, as labeled.}
\label{fig:xlfmod}
\end{figure}

We consider the two alternative models by CAK and Vink \etal, 
which were introduced in Sec.\ref{sec:windmodel}. XLFs computed for these
models are shown in Fig.\ref{fig:xlfmod}. It can be seen that the XLFs
overlap in the low-luminosity regime for all models, which means that the 
XLF is independent of the exact form of $f(M_c)$ (see above) in this 
region. The XLF's differential slope is close to -1 in this regime.
The position of the cross-over luminosity $L_{cr}$ is different for the 
three models with similar power-law exponents above respective $L_{cr}$.
The CAK model cuts off rather abruptly at luminosities
considerably below the Eddington luminosity of neutron stars and so 
appears unable to account for the observations. Barring the
difference in the $L_{cr}$, the Kudritzki-Reimers model and the 
Vink model give similar results, in general agreement
with observations.

\subsubsection{Effects of the IMF slope}
\label{sec:IMFvary}

The IMF is typically given as a single power-law in the mass range that is 
relevant to the problem at hand. The power-law index of the IMF ($\alpha$)
is considered to be within the range $\sim 2.0 - 2.7$, $\alpha = 2.35$
being the standard value given by Salpeter and widely used in this mass range. We
study the effect of varying $\alpha$ on XLF in the range 2.0 - 3.0. The
effect is shown in the Fig.\ref{fig:xlfmod}. The XLF at low luminosities
below the kink shows hardly any variation. We therefore display the XLF 
only for luminosities above the kink. Even above the kink, the XLF slope
varies by a small amount (from -1.47 to -1.66 between the lower and upper
limits to $\alpha$ considered here), showing relative insensitivity of the
XLF to the IMF slope. For an exact matching with the observed XLF, a 
slightly steeper IMF seems to be preferred, if other
parameters remain unchanged.

\subsubsection{Other effects}
\label{sec:othervary}

Among other parameters in the problem, we consider the stellar-wind velocity
$V_w$ and the metallicity $z$. The terminal velocity of the wind is an
important factor in determining the capture fraction. 
It is typically of the order of the
escape velocity at the surface of the companion and generally thought to
be $\sim 10^3$ km/s. We varied $V_w$ around this canonical value to
test the effect on the XLF. The changes in both the XLF slope and the
high-luminosity cutoff were insignificant.

\begin{figure}
\includegraphics[scale=1.0]{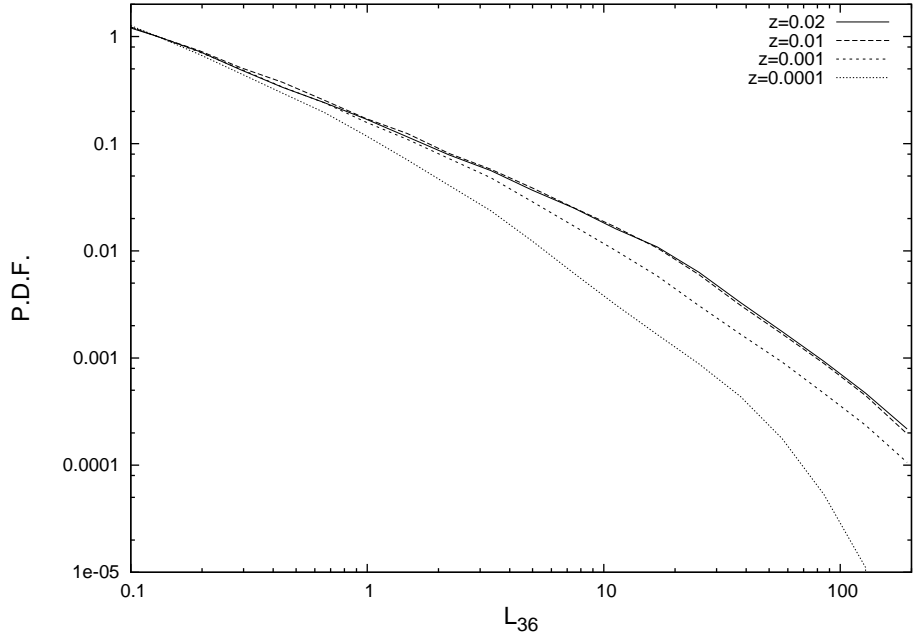}
\caption{Dependence of XLF on model parameters: metallicity $z$, as labeled.}
\label{fig:xlfmod1}
\end{figure}

Consider next the metallicity $z$, which affects the stellar parameters 
and the wind mass loss rate. We considered a large range of $z$
from $10^{-4}$ to $0.02$, and used the model of Vink \etal~(2001) to
study the effects of varying $z$ on our final results. The results are
shown in Fig.\ref{fig:xlfmod1}. 
There is no significant difference between the XLFs
for $z = 0.01$ and $z = 0.02$: both have a slope $\approx -1.6$ in the 
luminosity range $2\times 10^{37} - 2\times 10^{38}$ erg s$^{-1}$. The 
XLF for $z = 0.001$ also has nearly the same slope, but a somewhat 
different crossover luminosity. Finally, the XLF for $z = 10^{-4}$ is 
markedly different, but such low metallicities are not realistic for 
HMXBs, as they would be expected only in old stellar systems.

\section{Discussion}
\label{sec:discussion}

In this work, we have developed a strightforward, first-principles 
scheme for understanding collective properties of HMXB populations, 
starting from standard, well-known 
collective properties of primordial binaries
which are the progenitors of HMXBs, and following the transformations
of the probability distributions through the evolutionary processes that
lead to the formation of HMXBs. Our purpose, of course, was to assess
if the standard picture of primordial binaries and the standard 
evolutionary scenario for HMXBs together can account for the observed
collective properties of HMXBs in a basic, simple way. The fact that
we find that such an account can indeed be given is most encouraging, 
and it constitutes, in our view, an essential
step towards attempts at elaborate population synthesis schemes
designed for understanding further details of these collective
properties. Since our procedure is transparent and readily
understandable at each step, it is easy in our scheme to follow 
the role of each ingredient in shaping the final HMXB distribution.
In this section, we discuss various issues of principle and procedure      
which are relevant to this line of approach, and conclude with plans
for the future.

\subsection{Primordial-binary mass distribution}
\label{sec:mdist}

We pointed out in Sec.\ref{sec:primordial}
that distributions of the masses 
$M_p,M_s$ of the primary and secondary in the primordial
binary can be described either in terms of these masses 
themselves, \ie, the pair  ($M_p,M_s$), or alternatively in terms 
of the primary mass and the mass ratio $q\equiv M_s/M_p$, \ie,
the pair ($M_p,q$). Both have been done in the literature 
\citep{Warn,JasFer,Halb,Trim,Kouetal,KobFry}, 
and we have chosen here the second description for its closer
correspondence with essentially all recent work.
The relation between these two approaches has been discussed
thoroughly by Tout (1991). 
 
\subsection{Be-star binaries}
\label{sec:bestar}

Massive companions in HMXBs are of two types in general, \viz, OB 
giants/supergiants and Be stars, the latter being characterized by 
(a) strong, broad emission lines that supply evidence for rapid 
stellar rotation, and (b) somewhat lower masses and wider orbits. 
Since we have confined our detailed calculations in this work to 
the former, as stated in Sec.\ref{sec:psntoxrb}, we now discuss 
the expected role of Be-star binaries in HMXB XLF.  

It is believed
that Be stars are often surrounded by an outflowing disk of
matter expelled by centrifugal forces from the fast-roating equatorial
regions of the star, in addition to the usual stellar wind emitted
from all over the stellar surface. Accretion by the neutron star from
this outflow material generally follows the basic description from
stellar winds given in Sec.\ref{sec:windmodel}, with 
appropriate values of $V_w$ for 
the fast wind and the slowly outflowing disk, with one major 
caveat. Since the matter in the rapidly-rotating outflow disk has 
much angular momentum, it may form an {\it accretion} disk around the
neutron star, which would then drain on the neutron star a slow,
viscous timescale. This complicates the description considerably,
as the outflow disk is generally expected to be tilted with respect
to the orbital plane \citep{Ghosh95}, so that the orbiting neutron star would 
``crash'' through this outflow disk (twice per orbit in general),
possibly acquire an accretion disk, and accrete it slowly over the
rest of the orbit. This would naturally lead to outbursts of X-ray
emission, which are indeed observed in Be-star HMXBs.

The question for our purposes here is: how does all this affect the 
XLF of HMXBs? First consider the observational situation. 
The point to note here is that Be-star HMXBs 
are basically transient systems with low duty cycles,
so that a long-term average of the luminosity of a Be-system is 
much lower than that of an OB supergiant-system. Thus, in an XLF
constructed from a long-term monitoring of the X-ray sky with an
all-sky monitor, one would expect the high-$L$ parts to be 
dominated by OB-systems, while the low-$L$ parts may have 
considerable contributions from Be-systems. Indeed, the observed
XLFs cited earlier in this paper
have been constructed from recent work with
X-ray observatories in the following way. The XLF for the 
Milky Way has be constructed from $\sim 4$ years of observation
with the all-sky monitor on $RXTE$. Thus, the above argument
certainly applies to this case. Indeed, since the typical 
long-term average expected from Be-systems would be at or below
the lower end of the luminosity-range over which the XLF is
actually reported (and this applies to basically all observed XLFs 
except those obtained from $XMM-Newton$ observations of SMC and
LMC; see Sec.\ref{sec:compxlf}), we would expect little contribution 
to the reported XLF from Be-star systems. 

However, note that the XLF for other nearby galaxies reported in 
these references have been constructed essentially from one 
``snapshot'' (\ie, a single exposure) taken by $Chandra$ and 
$XMM-Newton$ \citep{Gil04}. 
Given this, it is remarkable that the XLFs of
all these galaxies (suitably normalized by their SFRs) are 
essentially coincident with one another over the same range of
$L$, as explained earlier. Considering the fact that, 
because of their low duty cycles, only a small fraction of the 
Be-systems would be present in these ``snapshot'' XLFs, we would
still expect some contribution from them in a range of $L$ typical
of the outbursts of Be-systems. The fact that these XLFs appear
very similar to the above long-term average XLF of the Milky Way 
over the canonical luminosity range $3\times 10^{35}{\rm erg s}
^{-1}\le L\le 3\times 10^{40}{\rm erg s}^{-1}$ is therefore most 
noteworthy, and may imply that, for reasons which are not 
clear at present, Be-systems may not have made a substantial 
contribution to these observed XLFs.

Now consider the inclusion of Be-systems in calculational schemes
like ours. To the extent that the wind-accretion formalism can
be applied to accretion from both the fast, low-density wind 
from the stellar surface and the slow, high-density outflow in
the equatorial disk \citep{Ghosh95}, these systems are already in 
our scheme, at least in principle.
However, while a quantitative description of the periodic 
acquisition and drainage of accretion disks described above has
been done for individual binary systems \citep{PraGho}, its 
inclusion in a study of collective properties of HMXB populations 
is more complex and outside the scope of this work.
        
\subsection{Black-hole HMXBs and ULXs}
\label{sec:blackhole}

As mentioned in Sec.\ref{sec:compxlf}, 
the observed ``universal'' XLF of HMXBs extends 
to about two decades of luminosity above the Eddington luminosity for 
canonical 1.4\Msun~neutron stars. While we focus on neutron-star HMXBs
in this study, it is interesting to consider this brightest end of the
XLF briefly. If the X-ray sources here are accretion-powered, they can 
only contain accreting black holes (some sources upto luminosities
$\sim 10^{39}$ erg s$^{-1}$ can of course be close juxtapositions of
several neutron-star sources, as has sometimes been suggested), either 
stellar-mass ones ($M_{BH}\sim 10\Msun$), or even the so-called
intermediate-mass black holes (IMBH), with masses $M_{BH}\sim 
(10^2-10^5)\Msun$, the X-ray sources corresponding to the latter 
objects being often called ultra-luminous X-ray sources (ULX).  
  
A truly remarkable feature of the 
universal HMXB XLF is that a single, smooth power law gives 
an excellent account of X-ray binaries containing neutron stars,
stellar-mass black holes, and IMBHs over the entire luminosity
range $3\times 10^{35}{\rm erg s}^{-1}\le L\le 3\times 10^{40}
{\rm erg s}^{-1}$, \ie, above the ``kink'' in the XLF \citep{Gil04}. 
While it is not difficult to extend 
the formation and evolution scenario
outlined in the earlier sections to include higher primary masses 
which would produce stellar-mass black holes, and then imagine 
that the other systematics would go through in such a way as to 
extend the power-law XLF, this argument does not automatically
include ULXs and IMBHs, whose formation scenario has to be
different and more exotic, \eg, black-hole mergers in dense stellar 
clusters. And yet, as Gilfanov (2004) has noted, these ``rare'' and 
``exotic'' objects appear to form a smooth extension of the ``ordinary'' 
HMXB population. Whether this is really true or not can possibly 
be probed with future observations in a way suggested by this author:
if we adopt the alternative hypothesis that the apparent cutoff in the 
currently observed XLF at $L\sim 3\times 10^{40}
{\rm erg s}^{-1}$ really corresponds to the maximum possible 
luminosity of what we may call ``ordinary'', stellar-mass black holes
referred to above, then these ``exotic'' IMBHs may show up beyond
this cutoff if we can observe regions with extremely high
star-formation rates, since the merger scenario for the formation of 
IMBHs implies that they would occur in very dense regions with 
very high star-formation rates. However, since such IMBHs would 
necessarily be rarer than stellar-mass black holes, we should expect 
a ``step down'' at the presently observed cutoff, beyond which the 
XLF would continue at a lower strength. This is a fascinating 
possibility. 

\subsection{Shape of the XLF}
\label{sec:xlfshape}

Understanding the XLF shape is an important step towards extracting 
information about various processes which determine the 
collective properties of HMXBs. Two main features of the XLF calculated
using our scheme are (1) A kink at $L_{cr} \approx 8 \times 10^{34}$ erg/s, 
and (2) a power-law behavior with the differential slope of -1.6 above 
$L_{cr}$. We will discuss the possible origin of these two features in 
this section.

\subsubsection{The XLF kink}
\label{sec:xlfkink}

\begin{figure}
\includegraphics[scale=0.6,angle=270]{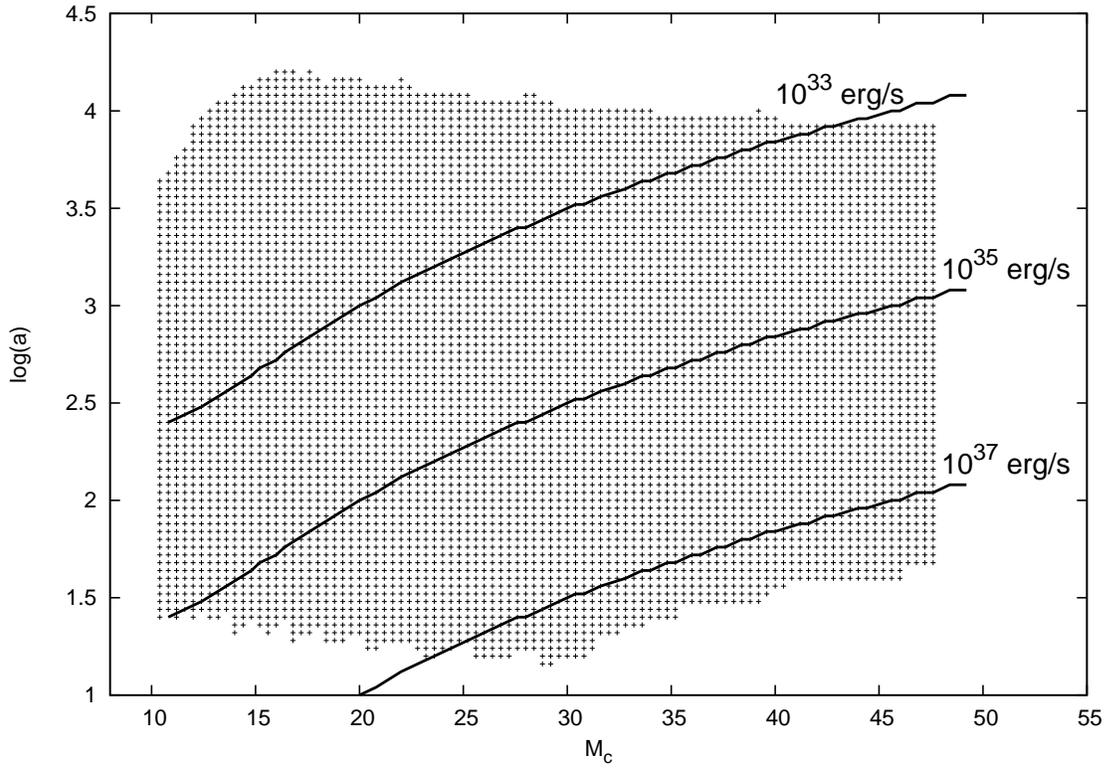}
\caption{Allowed zone in $M_c-a$ parameter space, shown as 
            shaded area. Overplotted are contours of constant L, 
            each contour labeled by its value of L.}
\label{fig:mcarange}
\end{figure}

The origin of the kink in the HMXB XLF can be understood in terms of
allowed zones in the parameter space. This is demonstrated in a clear
way by overplotting contours of constant $L$ on the allowed zone in
the $(M_c, a)$ plane, as shown in Fig.\ref{fig:mcarange}. Let us
first examine a simple ``toy'' model for the XLF.  In this model, we
approximate the post-SN PDF as a function of $M_c$ and $a$ as $f_{psn} 
\propto M_c^{-\alpha}/a$ (as discussed earlier, 
this ``toy'' is not too bad an approximation over much 
of the parameter range in case of no SN-kicks). We now note 
that $L$ can be written in a schematic way as $L\propto f(M_c)/a^2$. Next 
we transform the approximate form of the post-SN PDF from $(M_c, a)$ to 
$(M_c, L)$. A bit of simple algebra shows that such a transformation gives 
a PDF of the form $f_H\propto M_c^{-\alpha}/L$. Integrating this PDF 
over $M_c$ would yield $dN/dL\propto  L^{-1}$, if the limits of 
integration were independent of $L$.
Figure \ref{fig:mcarange} shows that this is indeed the case for
$10^{33} {\rm erg/s} \leq L \leq L_{cr}$, since the $M_c$-limits are 
clearly seen to be almost independent of $L$ when $L$ lies in 
this range.

By contrast, as we move to higher $L$-values,
the contours become shorter by cutting off low-$M_c$ regions. In
other words, the lower limit of integration becomes a function
of $L$. Shortening of the contour length results in a power-law XLF
with the slope steeper than -1. Eventually, near the neutron-star
Eddington luminosity, the contour passes out of the allowed region 
altogether, and the XLF is cut off.

At very low luminosities we expect a similar mechanism to cause a 
turnover in the XLF (which would cut it off eventually at extremely
low luminosities), as is clear from Fig.\ref{fig:mcarange}, since
higher-$M_c$ regions are progressively cut off as $L$ decreases in
this range. However, this effect is not expected to be 
very important for two reasons. First, cutting of the highest-mass 
regime in $M_c$ is not a severe problem, since the $M_c$-distribution
drops steeply in that regime anyway. Therefore, one
needs to go to very low luminosities ($\leq 10^{32}$ erg/s) to
observe this effect. Second, with current observational sensitivity,
we can track the XLF behavior below $L_{cr}$ only for a few nearby
galaxies, and the very low luminosities indicated above are not
even approached. It appears, therefore, that this XLF turnover is unlikely
to be amenable to observation in the near future, and accordingly we
do not consider it any further.

\subsubsection{The XLF slope}
\label{sec:xlfslope}

The XLF of HMXBs obtained observationally does not reach luminosities
below $L_{cr}$ for most of the galaxies (see Sec.\ref{sec:observe}).
Therefore, only the power-law regime above $L_{cr}$ can be compared
with observations in most cases. Physical origins of this
power-law index are an important aspect of our understanding of 
the collective properties of HMXBs. Note first that our calculation of the 
XLF is a numerical one, and the power-law result given in Sec.
\ref{sec:hxlf} is only an analytic approximation
to it. However, simple qualitative arguments may serve to illustrate
the basic physics underlying such calculations, and we consider such
arguments below.

In a pioneering argument of this type, Postnov (2003;
also see Postnov \& Kuranov 2005) proceeded as follows. 
Expressing the XLF as:

\begin{equation}
{dN\over dL} = {dN\over dM_c}{dM_c\over dL},
\label{eqn:postnov}
\end{equation}     

one can use a suitable estimate of the mass-function $dN/dM_c$ 
of the companion for these arguments.  The estimation of 
$dM_c\over dL$ is more involved, and the original Postnov method
was to estimate $dL\over dM_c$ as follows, and use its reciprocal.
Using $L \propto \dot{M} \propto\dot{M}_w/(a^2V_w^4)$ 
(see Sec.\ref{sec:psntoxrb}),
this author obtained a power-law form $L\propto M_c^{\beta}$, with
the aid of a simple stellar-wind model $\dot{M}_w\propto L_c/V_w
\propto L_c\sqrt{R_c/M_c}$, into which the following scalings for
massive stars were inserted: $L_c\propto M_c^3$, $R_c\propto 
M_c^{0.8}$. This gave $\beta\approx 2.5$. Finally, assuming a
power-law form for the companion mass distribution, \ie, 
$dN/dM_c\propto M_c^{-\alpha}$, straightforward algebra with
Eq.(\ref{eqn:postnov})  leads to an XLF slope of $dN/dL\propto 
L^{-(\alpha+\beta -1)/\beta}$. Postnov (2003) assumed $\alpha = 2.35$ (\ie,
the exponent of the Salpeter IMF) for the $M_C$-distribution, 
which led to an XLF slope of $\approx -1.54$.  

In revisiting the above argument, we note first that Eq.(\ref{eqn:postnov})  
is incomplete, since N is a function of both $M_c$ and $a$, whose form
we have calculated explicitly in the earlier sections. Thus, the complete 
equation is

\begin{equation}
{dN\over dL} = {\partial N\over\partial M_c}{dM_c\over dL} +
{\partial N\over\partial a}{da\over dL}~.
\label{eqn:bhadghosh}
\end{equation}
 
Next, we note that the scalings of $L_c$ and $R_c$ with $M_c$ 
used in the Postnov (2003) work apply to massive {\it main-sequence} 
stars, but not to the evolved massive stars 
of interest here. Also, the wind mass-loss prescription used in 
that work is similar to that in the CAK model of the 1970s, 
which our calculations have shown to be inadequate. The rough 
scalings for $L_c$ and $R_c$ for evolved massive companions of 
interest here can be obtained from the appropriate formulae in 
Sec.5 of HPT, and are $L_c\propto M_c^{1.8}$, 
$R_c\propto M_c^{4.6}$.

With these scalings, and our $M_c$-distribution shown in 
Fig.\ref{fig:mcdf}, which we can approximate with
the power-law of exponent $\alpha = 2.4$ that applies to its 
principal part, we can evaluate the first term on the right-hand side
of Eq.(\ref{eqn:bhadghosh}). For the Kudritzki-Reimers model, 
it can be easily shown that $\beta\approx 5.4$, so that
this first term gives an XLF slope $-(\alpha+\beta -1)/\beta\approx -1.27$.     
The other stellar model gives an essentially identical final
result. The rest of the contribution comes from the second term 
on the right-hand side of Eq.(\ref{eqn:bhadghosh}), which can be easily
calculated and which leads to our overall XLF slope $\approx -1.63$.

We have gone through this argument in detail because it addresses an
interesting observation made by Gilfanov (2004) that, in HMXB 
systems powered by stellar-wind accretion, the distribution of $\dot{M}$
and therefore $L$ should be governed by the properties of the massive
companion, in particular the distribution of $M_c$ and $L_c$. Our 
work here shows that this is largely, but not completely, true. The 
properies of the massive companion are contained in the first term 
on the right-hand side of Eq.(\ref{eqn:bhadghosh}), and keeping only
this term, as Postnov (2003) did, amounts to neglecting altogether
the binary orbital properties which are contained in the second term 
on the right-hand side of Eq.(\ref{eqn:bhadghosh}), and which also 
influence the XLF. The estimates summarized above give a measure of
the relative sizes of the effects of the companion and the orbit, and 
demonstrate that, while the latter are certainly smaller, they are by 
no means negligible.     
   
\subsection{Conclusions and outlook}
\label{sec:conclude}

In this work, we have described a method for obtaining the
distributions of some of the essential collective properties of HMXB
populations in the stellar fields of normal/starburst galaxies,
wherein we start from accepted distributions of primordial binaries
which are progenitors of such HMXBs, and follow the transformation 
of these distributions with the aid of a Jacobian formalism as the 
primordial binary population evolves into the HMXB population. 
Our method, which is semi-analytic, traces in a transparent way the
effects of various processes in the course of this evolution, and so 
assesses with ease which physical processes dominate in determining
which distribution. For example, the distribution of the properties of
the massive companions seems to have the dominant effect on the XLF, 
although the distribution of the orbital parameters does have a 
significant effect, as we demonstrated
in Sec.\ref{sec:xlfslope}. But the distribution of HMXB orbital
periods appears to be strongly influenced by both the primordial
orbital distribution and the SN-kick properties. 

The agreement between our calculated XLF and binary-period
distribution and the observed HMXB distributions is most encouraging, 
and it justifies a future Monte Carlo population synthesis scheme 
for a more detailed undrstanding of how HMXB populations are built
in the stellar fields of normal galaxies. However, we must first 
extend our present method to the more complex problem of following the 
formation and evolution of LMXB populations from their corresponding 
primordial binaries, as mentioned in Sec.\ref{sec:intro}. This will
occupy us in the next papers in this series.       


\acknowledgments

It is a pleasure to thank M. Gilfanov, E. P. J.van den Heuvel, V. Kalogera,
L. Stella, and R. A. Sunyaev for stimulating discussions, and an
anonymous referee for valuable comments which greatly improved the paper.



\appendix

\section{Averaging supernova kicks and post-SN transformation equations}
\label{app1}

\subsection{Calculation of SN-kick average}
\label{snkick}

In this appendix we describe the method of averaging the effects of
SN-kicks for a Maxwellian kick-distribution. It has been widely assumed
that the distribution of kicks will be isotropic and recent observations
by have supported it \citep{hobbs}. Therefore the angle dependence
in Eq.\ref{eqn:postsn} is averaged in a straightforward way, which
leads to Eq.\ref{eqn:postsnav}. Two points are to be noted
here. First, the only surviving kick-term in the averaged transformation equations 
is \vsqav\, which is to be obtained by averaging $v^2$ over the Maxwellian 
distribution. Second, it must be remembered that the process of averaging gives
the average values of post-SN quantities. These therefore should be compared
with the average behaviour of a number of sample observational sets.

A further, most crucial, point to be noted at this stage is that in
our study here (and in all similar studies), we are interested in \emph{only} 
those post-SN systems which remain bound as XRBs, so that they can 
eventually produce HMXBs whose distribution we are interested in. 
Accordingly, we must exclude at this point all systems which become 
unbound in the SN. The way we do so is as follows. We note that, in
the $v^2$-averaging process, we would make an error if we carried out
the integrations over the Maxwellians in Eq.\ref{eqn:vsqav} 
upto infinitely large values of $v$, since this would include all the 
unbound systems that are to be excluded. Rather, we must
\emph{truncate} the integration at a suitable upper limit $v_{up}$
which corresponds to the point at which the post-SN system becomes 
just unbound. This point is readily obtained from the first of
Eqns.\ref{eqn:postsnav} by setting $a_f\rightarrow\infty$,
and $v_{up}$ given by:

\begin{equation}
v_{up} = \sqrt{\frac{M_i^t}{a_i}\left(1 - \frac{2\Delta M}{M_i^t}\right)}
\label{eqn:vup}
\end{equation}

We thus work with a Maxwellian which is truncated from above at $v_{up}$
(and suitably normalised), so that \vsqav\ is given by: 

\begin{equation}
\vsqav = \frac{\int_0^{v_{up}}v^4 exp(-v^2/2\sigma^2)dv}
       {\int_0^{v_{up}}v^2 exp(-v^2/2\sigma^2)dv}
\label{eqn:vsqavup}
\end{equation}

$v_{up}$ is well-defined for each set of values of pre-SN parameters.
For algebraic convenience, we can express it as 
$v_{up} = f\sqrt{2}\sigma$ for a given Maxwellian distribution
with dispersion $\sigma$. This simplifies the expression for \vsqav, which
can be written as $\vsqav = \sigma^2 h(f)$, where $h(f)$ is given by:

\begin{eqnarray}
h(f) &=& 3 - \frac{2f^2h_2(f)}{h_1(f) -h_2(f)} \nonumber \\
h_1(f) &=& \sqrt{\frac{\pi}{2}} Erf(f) \nonumber \\
h_2(f) &=& f \sqrt{2} e^{-f^2}
\label{eqn:h1h2}
\end{eqnarray}

\begin{figure}
\includegraphics[scale=0.6,angle=270]{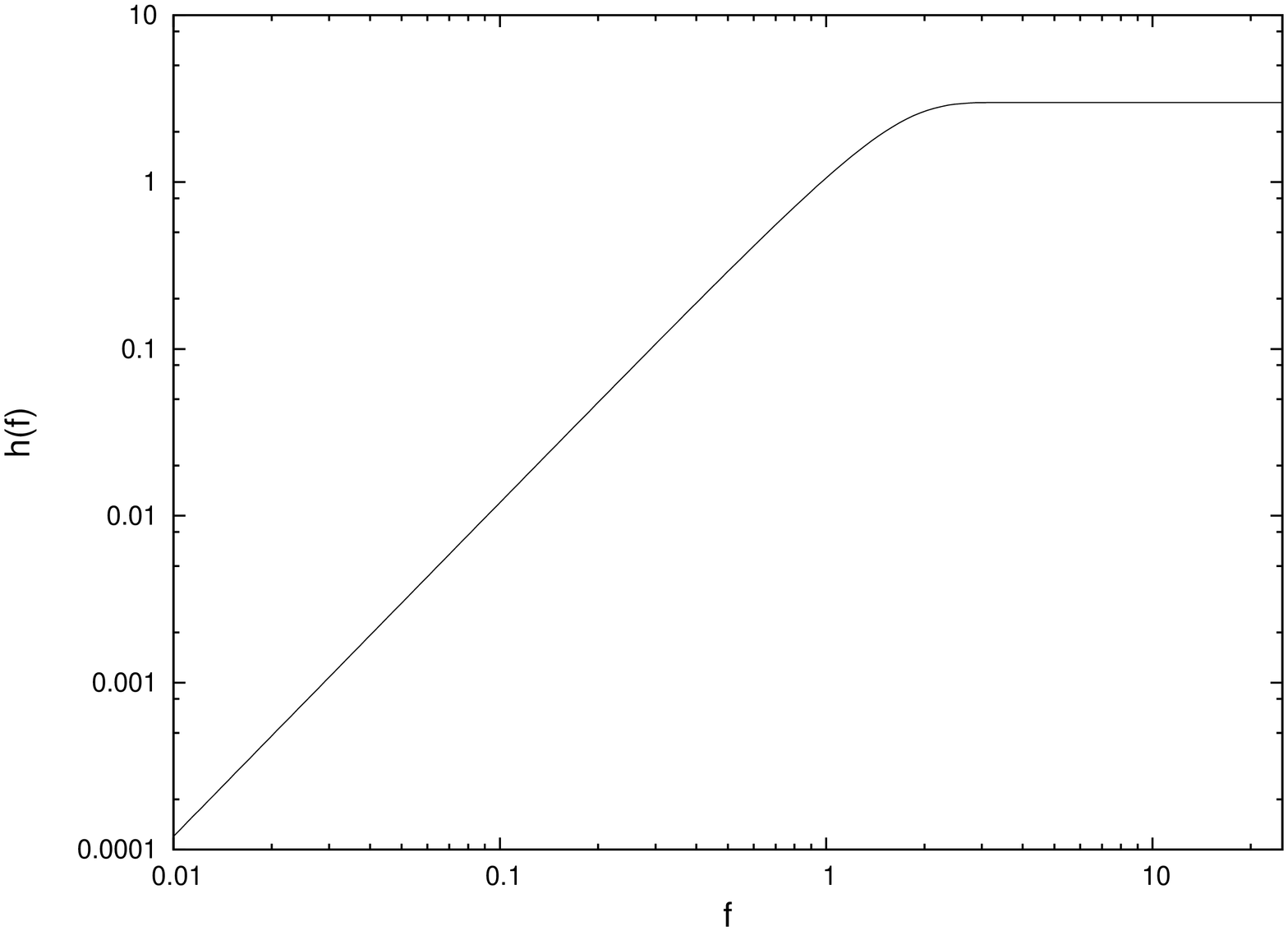}
\caption{Dependence of h(f) on f}
\label{fig:fhf}
\end{figure}

Fig.\ref{fig:fhf} shows $h(f)$ as a function of $f$. Two regimes are
clearly demarcated, with a small transition region between the two. The
low-$f$ region can be described by a power-law given by, $h(f) = 1.2f^2$,
whereas in the high-$f$ region, $h(f)=3$, which is its asymptotic value. The 
crossover point of the two regimes is at $f_c \approx \sqrt{2.5}$. This 
clear division in two regions give us a straightforward relation for \vsqav.
For $f<f_c$, \vsqav\ is independent of $\sigma$, and is given by $\vsqav=0.6v_{up}^2$.
On the other hand, for $f>f_c$ \vsqav\ is independent of $v_{up}$, and
is given by $\vsqav=3\sigma^2$. This behavior of $h(f)$ can be understood as follows.
For $f<f_c$, $v_{up}$ is small and hence only the initial rising part
of the Maxwellian is relevant. Thus the distribution is essentially given by 
$f(v) \propto v^2$, which is independent of $\sigma$. On the contrary, for
$f>f_c$, $v_{up}$ is large and almost the entire Maxwellian profile is
included, excluding only a small tail. Detailed calculations given
above show that the transition region between these two regimes is
small, so that, as a first approximation, we can use the
following simple prescription:

\begin{equation}
\vsqav = \left\{ \begin{array}{ll}
    0.6 v_{up}^2 & f < f_c \\
    3 \sigma^2 & f > f_c
    \end{array} \right. \;\;\;\; \mbox{where} f_c = \sqrt{2.5}
\label{eqn:vsqapprox}
\end{equation}

The above prescription can be substituted appropriately in eqn. \ref{eqn:postsnav}
to obtain the post-SN parameter transformation or the inverse transformation,
which is required for the Jacobian formalism. We describe in the next section
the relevant transformation equations and the Jacobian for the two cases.

\subsection{Post-SN parameter transformation}
\label{psntr}

\subsubsection*{For $f>f_c$}

In this region, \vsqav\ is independent of the parameters of the binary system.
We can simplify Eqn.\ref{eqn:postsnav} by using following ratios. Let
$r_i = M_i^t/a_i$, $r_f = M_f^t/a_f$, and $r_k = \sigma^2$. Also, we define
orbit size change factor as $s = a_f/a_i$. Equation \ref{eqn:postsnav}  
can then be used to write the inverse transformations:

\begin{eqnarray}
s &=& \frac{1+\sqrt{e^2\left(1+r_k/r_f\right) - r_k/r_f}}{1-e^2} \nonumber \\
r_i &=& r_f(2s-1) - 3r_k
\label{eqn:psninv1}
\end{eqnarray}

These equations are indirect transformations which give $a_i$ and $M_i^t$ on
substituting the definitions of $s$ and $r_i$ in above equations. Since the
companion mass is unchanged in the SN explosion, one can readily calculate
$M_{p,c}$ from $M_i^t$ for known $M_c$. The Jacobian of this transformation can
be calculated as follows. Note first that the transformation is only in two
variables \ie\ $(a_f, e) \rightarrow (a_i, M_{p,c})$, since the third 
parameter \ie\ $M_c$ is unchanged. For the same reason, one can also write
$\partial M_{p,c}/\partial (a,e) = \partial M_i^t/\partial(a,e)$. All the
necessary partial derivatives can be written in terms of the derivatives of
$s$ and $r_i$ as follows:

\begin{eqnarray}
\frac{\partial a_i}{\partial a_f} &=& \frac{1}{s} - \frac{a_f}{s^2}
\frac{\partial s}{\partial a_f} \nonumber \\
\frac{\partial a_i}{\partial e} &=& \frac{a_f}{s^2}
\frac{\partial s}{\partial e} \nonumber \\
\frac{\partial M_i^t}{\partial(a_f, e)} &=& 
a_i \frac{\partial r_i}{\partial(a_f, e)} + 
r_i \frac{\partial a_i}{\partial(a_f, e)} \nonumber \\
J_{sn} &=& \left| 
\frac{\partial M_{p,c}}{\partial e} \frac{\partial a_i}{\partial a_f} - 
\frac{\partial M_{p,c}}{\partial a_f} \frac{\partial a_i}{\partial e} \right|
\label{eqn:psnjac1}
\end{eqnarray}

\subsubsection*{For $f<f_c$}

\vsqav\ in this region is independent of $\sigma$ and written only in terms 
of $v_{up}$, which is given by eqn. \ref{eqn:vup}. The inverse transformations
in this case are given by:

\begin{eqnarray}
s &=& \frac{1+\sqrt{\left(3e^2 - 1\right)/2}}{1-e^2} \nonumber \\
M_i^t &=& \frac{M_f^t}{3}\left[5s\left(1-e^2\right) - 4\right]
\label{eqn:psninv2}
\end{eqnarray}

where $s$ is the orbital-size factor defined similarly as $a_f/a_i$. The
Jacobian in this case is given by an even simpler relation. One can easily see
that $M_i^t$ is independent of $a_f$ and depends only on e. Therefore
$\partial M_i^t/\partial a_f = 0$. The Jacobian in that case is given by
$J_{sn} = (\partial M_i^t/\partial e)(\partial a_i/\partial a_f)$. 
Straightforward algebra shows that 

\begin{equation}
J_{sn} = \frac{5}{2s} \frac{M_f^te}{s\left(1-e^2\right)+1}
\label{eqn:psnjac2}
\end{equation}

These two inverse transformations can be applied in the relevant regions
to transform the PDF with the aid of the Jacobian formalism.





\clearpage

\clearpage

\clearpage

\clearpage
\clearpage
\clearpage
\clearpage


\clearpage

\clearpage
\clearpage

\clearpage
\clearpage
\clearpage


\begin{thebibliography}{}

\bibitem[Arzoumanian et al. 2002]{arzoumanian}
Arzoumanian, Z., Chernoﬀ, D. F., and Cordes, J. M. 
2002, \apj, 568, 289 

\bibitem[Belczynski et al. 2008]{belc08}
Belczynski, K., Kalogera, V., Rasio, F. A., Taam, R. E., Zezas, A., Bulik, T.,
Maccarone, T. J., and Ivanova, N. 2008, ApJS, 174, 223 

\bibitem[]{BG06}
Banerjee, S., and Ghosh, P. 2006, \mnras, 373, 1188

\bibitem[]{BG07}
Banerjee, S., and Ghosh, P. 2007, \apj, 670, 1090

\bibitem[]{BG08}
Banerjee, S., and Ghosh, P. 2008, \apj, 680, 1438

\bibitem[Castor et al. 1975]{castor}
Castor, J. I., Abbott, D. C., and Klein, R. I. 1975, \apj, 195, 157.

\bibitem[Davidson \& Ostriker 1973]{DavOs}
Davidson, K. and Ostriker, J. P. 1973, \apj, 179, 585.

\bibitem[Ghosh 1995]{Ghosh95}
Ghosh, P. 1995, \apj, 453, 411

\bibitem[Ghosh 2007]{Ghosh}
Ghosh, P. 2007, {\it Rotation and Accretion Powered Pulsars},
World Scientific, Singapore.

\bibitem[Ghosh \& White 2001]{GW01}
Ghosh, P., and White, N. 2001, \apj, 559, L97

\bibitem[Giacconi et al. 1971]{giet71}
Giacconi, R., Gursky, H., Kellogg, E., Schreier, E. \& Tananbaum,
H. 1971, \apj, 167, L67.

\bibitem[Gilfanov 2004]{Gil04}
Gilfanov, M. 2004, Prog Theo Phys Suppl, 155, 49.

\bibitem[Gilfanov et al. 2004a]{GGS04a}
Gilfanov, M. , Grimm, H.-J., and Sunyaev, R. 2004a, 
Nucl Phys B Proc Suppl, 132, 369.

\bibitem[Gilfanov et al. 2004b]{GGS04b}
Gilfanov, M. , Grimm, H.-J., and Sunyaev, R. 2004b, 
\mnras, 347, L57.

\bibitem[Grimm et al. 2002]{Grimetal02} 
H.-J. Grimm, Gilfanov, M. , and Sunyeav 2002, \aap, 391, 923.

\bibitem[Grimm et al. 2003]{Grimetal03} 
H.-J. Grimm, Gilfanov, M. , and Sunyeav 2003
ChJAS, 3 (supplement), 257.

\bibitem[Halbwachs 1983]{Halb}
Halbwachs, J. L. 1983, \aap, 128, 399.

\bibitem[Heger et al. 2003]{Hegetal03}
Heger, A., et al., 2003. \apj, 591, 288

\bibitem[Hobbs et al. 2003]{hobbs}
Hobbs, G., Lorimer, D. R., Lyne, A. G., and 
Kramer, M. 2005, \mnras, 360, 974

\bibitem[Hurley et al. 2000]{HPT00}
Hurley, J., Pols, O., and Tout, C. 2000, \mnras, 315, 543

\bibitem[Jaschek \& Ferrer 1972]{JasFer}
Jaschek, C., and Ferrer, O. 1972, \pasp, 84, 292.

\bibitem[Kalogera 1996]{kalogera96a}
Kalogera, V. 1996, ApJ, 471, 352 

\bibitem[Kim \& Fabbiano 2004]{KimFab04}
Kim, D.-W., and Fabbiano, G., \apj, 611, 846

\bibitem[Kim \& Fabbiano 2010]{KimFab10}
Kim, D.-W., and Fabbiano, G., \apj, 721, 1523

\bibitem[Kobulnicky \& Fryer 2007]{KobFry}
Kobulnicky, H., and Fryer, C. 2007, \apj, 670, 747

\bibitem[Kouwenhoven et al. 2007]{Kouetal}
Kouwenhoven, M., et al. 2007, \aap, 474, 77

\bibitem[Kroupa \& Weidner 2003]{Kroup}
Kroupa, P. and Weidner, C. 2003, \apj, 598, 1076

\bibitem[Kudritzki \& Reimers 1978]{KudReim}
Kudritzki, R., and Reimers, D. 1978, \aap, 70, 227.

\bibitem[Lamb et al. 1973]{Lambet1}
Lamb, F. K., Pethick, C. J. and Pines, D. 1973, \apj, 184, 271.

\bibitem[Linden et al. 2009]{linden}
Linden, T., Sepinsky, J. F., Kalogera, V., and Belczynski, K. 2009, 
ApJ, 699, 1573 

\bibitem[Liu et al. 2005]{LPH05}
Liu, Q. Z., van Paradijs, J. and van den Heuvel, E. P.J. 2005, 
\aap, 442, 1135.

\bibitem[Liu et al. 2006]{LPH06}
Liu, Q. Z., van Paradijs, J. and van den Heuvel, E. P.J. 2006,
\aap, 455, 1165 

\bibitem[Madau et al. 1998]{Madau}
Madau, P., Pozzetti, L., and Dickinson, M. 1998, \mnras, 498, 106 

\bibitem[\"Opik 1924]{Opik}
\"Opik, E. 1924, Tartu Obs Publ, 25, No. 6.

\bibitem[Podsiadlowski et al. 2004]{pod04}
Podsiadlowski, P., Langer, N., Poelarends, A. J. T., 
Rappaport, S., Heger, A., and Pfahl, E. 2004, ApJ, 612, 1044

\bibitem[Postnov 2003]{Post}
Postnov, K. 2003, Astr Let, 29, 372

\bibitem[Postnov \& Kuranov 2005]{PosKur}
Postnov, K.,and Kuranov A. 2005, Astr Let, 31, 7

\bibitem[Pravdo \& Ghosh 2001]{PraGho}
Pravdo, S., and Ghosh, P. 2001, \apj, 554, 383

\bibitem[Pringle \& Rees 1972]{PriRe}
Pringle, J. E. and Rees, M. J.  1972, \aap, 21, 1.

\bibitem[Reimers 1975]{reimers}
Reimers, D. 1975, Mem Soc Roy Sci Li\`ege, 6e serie, 8, 369.

\bibitem[Salpeter 1955]{Salp}
Salpeter, E. E. 1955, \apj, 121, 161

\bibitem[Sana \& Evans 2011]{SanEva}
Sana, H., and Evans, C. 2011, in {\it Active OB stars}, Proc IAU
Symp 272, eds. C. Neiner et al., in press.

\bibitem[Scheck et al. 2006]{scheck06}
Scheck, L., Kifonidis, K., Janka, H.-T., and 
M\"uller, E. 2006, \aap, 457, 963 

\bibitem[Scheck et al. 2004]{scheck04}
Scheck, L., Plewa, T., Janka, H.-T., Kifonidis, K., 
and M\"uller, E. 2004, Phys. Rev. Let., 92, 011103

\bibitem[Shapiro \& Teukolsky 1983]{ShaTu}
Shapiro, S. L. and Teukolsky, S. A.  1983, {\it Black Holes,
White Dwarfs, and Neutron Stars: The Physics of Compact Objects},
Wiley \& Sons, New York.

\bibitem[Shtykovskiy \& Gilfanov 2005a]{SG05a}
Shtykovskiy, P. and Gilfanov, M. 2005a, \mnras, 362, 879.

\bibitem[Shtykovskiy \& Gilfanov 2005b]{SG05b}
Shtykovskiy, P. and Gilfanov, M. 2005b, \aap, 431, 597.

\bibitem[Tout 1991]{Tout}
Tout, C. 1991, \mnras, 250, 701.

\bibitem[Trimble 1990]{Trim}
Trimble, V. 1990, \mnras, 242, 79.

\bibitem[van den Heuvel 1983]{vdH83}
van den Heuvel, E. P. J. 1983, in {\it Accretion-driven stellar X-ray sources}, 
eds. W. H. G. Lewin and E. P. J. van den Heuvel, Cambridge Univ. 
Press, Cambridge, p.~303.

\bibitem[van den Heuvel 1991]{vdH91}
van den Heuvel, E. P. J. 1991, in {\it Neutron stars: theory and observation}, 
eds. J. Ventura and D. Pines, Kluwer, Dordrecht, p.~171. 

\bibitem[van den Heuvel 1992]{vdH92}
van den Heuvel, E. P. J. 1992, in {\it X-ray binaries and recycled pulsars}, 
eds. E. P. J. van den Heuvel and S. A. Rappaport, Kluwer,  Dordrecht, p.~233.  

\bibitem[van den Heuvel 2001]{vdH01}
van den Heuvel, E. P. J. 2001, in {\it The neutron star-black hole 
connection}, eds. C. Kouveliotou et al., Kluwer, Dordrecht, p.~173. 

\bibitem[Vink et al. 2000]{vink}
Vink, J. S., de Koter, A., and Lamers, H. J. G. L. M. 
2000, \aap, 362, 295.

\bibitem[Vink et al. 2001]{vink01}
Vink, J. S., de Koter, A., and Lamers, H. J. G. L. M. 
2001, \aap, 369, 574. 

\bibitem[Warner 1961]{Warn}
Warner, B. 1961, \pasp, 73, 439. 
 
\end{thebibliography}
\end{document}